\documentclass[11pt]{iopart} 
\usepackage{graphicx}
\usepackage{iopams}
\usepackage{setstack}
\begin{document}
\jl{1}
\title[Continued-fraction representation of the Kraus map for reservoir damping]{Continued-fraction representation of the Kraus map for non-Markovian reservoir damping}

\author{A J van Wonderen and L G Suttorp}
\address{Institute of Physics, University of  
Amsterdam, \\ Science Park 904, NL-1098 XH Amsterdam, The Netherlands}
\ead{vanwonderen@planet.nl}

\begin{abstract} 
Quantum dissipation is studied for a discrete system that linearly interacts with a reservoir of harmonic oscillators at thermal equilibrium. 
Initial correlations between system and reservoir are assumed to be absent. The dissipative dynamics as determined by the unitary evolution 
of system and reservoir is described by a Kraus map consisting of an infinite number of matrices. For all Laplace-transformed Kraus matrices 
exact solutions are constructed in terms of continued fractions that depend on the pair correlation functions of the reservoir. By performing 
factorizations in the Kraus map a perturbation theory is set up that conserves in arbitrary perturbative order both positivity and probability 
of the density matrix. The latter is determined by an integral equation for a bitemporal matrix and a finite hierarchy for Kraus matrices. 
In lowest perturbative order this hierarchy reduces to one equation for one Kraus matrix. Its solution is given by a continued fraction of a 
much simpler structure as compared to the non-perturbative case. In lowest perturbative order our non-Markovian evolution equations are applied 
to the damped Jaynes-Cummings model. From the solution for the atomic density matrix it is found that the atom may remain in the state of 
maximum entropy for a significant time span that depends on the initial energy of the radiation field.  
\end{abstract}

\pacs{03.65.-w, 05.30.-d, 42.50.-p} 

\vspace{8mm}  

\noindent{\it Keywords: quantum dissipation, non-Markovian dynamics, Kraus maps}

\submitted 

\maketitle 

\section{Introduction} 

The Wigner-Weisskopf theory of natural linewidth \cite{WW:1930} is one of the earliest descriptions of a quantum system that 
exchanges energy with an electromagnetic radiation field. The role of the open system is played by a two-level atom which 
spontaneously emits a photon by making a transition from the excited state $|2\rangle $ of energy $\omega_{2}$ to the ground state 
$|1\rangle $ of energy $\omega_{1}$. The atomic decay is modeled with the help of an exponential function. Furthermore, in the 
interaction Hamiltonian between atom and electromagnetic field counter-rotating terms are discarded. Owing to these simplifications, 
the lifetime of the excited atomic state can be readily expressed in terms of the coupling constant $g(\omega)$ between atom and field 
mode of frequency $\omega$. A mathematically rigorous treatment of the Wigner-Weisskopf atom is furnished in \cite{JAK:2006}. 

For times smaller than $|g(\omega_{2}-\omega_{1})|^{-2}$ the behaviour of the two-level atom is substantially influenced by transient 
effects. Consequently, the atomic evolution in time $t$ cannot be described on the basis of exponential functions. Hence, the atomic 
density matrix $\rho_a(t)$ cannot be written as a Markovian map $\mbox{exp} (Lt)\rho_a(0)$, with a generator $L$ independent of time. 
Indeed, still dropping counter-rotating terms one determines the lower diagonal element of the atomic density matrix for arbitrary 
times as   
\[   
\hspace{-20mm} 
\langle 2| \rho_a (t) |2 \rangle = 
\left|\, \int_{-\infty}^{\infty} \frac{\mbox{d}\omega}{2\pi} \exp{(-i\omega t)} 
\left[\omega - \int_0^{\infty} \mbox{d} \omega' 
\frac{|g(\omega')|^2}{\omega + \omega_{2} -\omega_{1} - \omega' + i0} \right]^{-1}\, \right|^2 \, ,   \nonumber 
\]
\begin{equation}  
\label{1} 
\end{equation} 
where the purely imaginary number $i0$ lies above and infinitesimally close to the real axis. The standard result (\ref{1}), which is valid 
at zero temperature of the radiation field, has been known long since and has been featured in textbooks \cite{KNI:1972, LOU:1973}. 

Despite its advanced age extensions of (\ref{1}) are scarce. In \cite{DAV:1974} radiative decay of a multi-level atom was studied. However, 
the assumption of weak coupling between atom and electromagnetic field was made. Then the dissipative dynamics can be generated from a 
Markovian map. If one stays outside the weak-coupling limit, a dynamical map with constant generator $L$ does not exist. For this regime 
extensions of (\ref{1}) to such relevant cases as an electromagnetic continuum of finite temperature or a decaying multi-level atom have 
not been published in the literature. The present work aims at filling up these gaps in our knowledge of quantum dynamics. 

In realizing our goal, previous findings  might give us a clue as to what shape extensions of (\ref{1}) should possess. More specifically, 
in \cite{JPA:2006} dissipative quantum evolution was studied in a separable Hilbert space that was coupled to a reservoir of zero 
temperature through an interaction free of counter-rotating terms. For this simplified setting continued fractions made their appearance 
in the density matrix of the open system. Moreover, the solution for the density matrix was found to reproduce (\ref{1}) upon choosing a 
Hilbert space of dimension two. We thus may surmise that continued fractions will constitute an effective tool in expanding on the 
fundamental formula (\ref{1}). Previous proposals for studying quantum dissipation outside the weak-coupling limit include employment 
of the Feshbach projection method \cite{CHR:2013} and use of one-dimensional projection operators \cite{SEM:2016}. The derivation of 
exact results outside the weak-coupling limit is a problem of long standing \cite{HAA:1969, HAS:1977} and has turned out to be a very 
hard task \cite{FER:2017}. 

Inspired by the foregoing considerations, we are going to devise a finite-temperature theory of non-Markovian quantum dynamics on the basis of 
matrix continued fractions. Our treatment is founded on the following three assumptions: (i) the Hilbert space of the open quantum system is 
separable, i.e., it is spanned by a countable number of ket vectors; (ii) initial correlations between the open quantum system and the 
surrounding reservoir are absent, so that the initial composite density operator for system and reservoir factorizes; in reality, initial 
correlations may influence the dissipative dynamics significantly \cite{SHA:2009}--\cite{CHI:2015}; (iii) the reservoir consists of a continuum 
of harmonic oscillators that are initially at thermal equilibrium and that linearly interact with the system potentials. This choice is 
indispensable to our treatment, because it allows us to  factorize all higher-order correlation functions of the reservoir with the help of 
Wick's theorem. Thus the influence of the reservoir is completely accounted for by a set of pair correlation functions.  

Adopting the setting (i)-(iii), we show in section 2 that an extension of the Markovian map $\exp ( L t ) \rho_S$ exists that is valid outside 
the regime of weak coupling. This non-Markovian map takes on the form of $ {\cal T} \exp [ L(t) ]\rho_S$, with $ {\cal T}$ a time-ordering 
prescription and $\rho_S$ the initial state of the open system. The pair correlation functions of the reservoir enter through the 
time-dependent generator $L(t)$. In the limit of weak coupling the latter reduces to the standard generator that is known from the 
literature \cite{DAV:1973}. Upon carefully carrying out the time-ordering prescription, our non-Markovian map reduces to an infinite expansion 
in terms of Kraus matrices \cite{KRA:1971}. 

Next, by suitably enlarging the set of Kraus matrices and performing an integral transformation of Laplace type, a closed hierarchy of 
nonlinear equations is constructed. For reason of clarity, this job is divided into two parts. In subsection 3.1 all of the necessary 
definitions and notations are gathered, while in subsection 3.2 the derivation of the nonlinear hierarchy is carried out. Its iteration 
yields exact solutions for the Kraus matrices in terms of infinite continued fractions. The actual execution of this last iterative process 
is technically demanding and therefore deferred to another paper \cite{arXiv:2017}.     
   
In section 4, we attempt to reduce the technical barriers to our formalism by adopting a perturbative approach. Truncating the infinite 
set of Kraus matrices and factorizing Kraus matrices of any order exceeding the truncation parameter, we arrive at a finite set of integral 
equations for modelling non-Markovian evolution of an open quantum system. The perturbative density matrix complies with conservation of 
both positivity and probability, for any value of the truncation parameter. 

In lowest perturbative order, the complete dissipative dynamics can be described in terms of one Kraus matrix only. The solution for the 
latter is given by a continued fraction possessing a much simpler structure than its counterpart of the exact case. Therefore, applications 
of the theory can be worked out analytically now. This is demonstrated in section 5, where we find the non-Markovian density matrix of the 
resonant Jaynes-Cummings model. We introduce radiative damping by coupling the two-level atom to a transverse electromagnetic continuum of 
zero temperature through an interaction that does not contain counter-rotating contributions.   

In addition to the setting (i)-(iii), we furthermore assume that all of the infinite continued fractions occurring in this paper are 
convergent. We recall that the use of continued fractions in quantum optics goes back to early work on the Rabi model by Schweber 
\cite{SCH:1967} and Swain \cite{SWA:1973}. A few years ago, these treatments have been reviewed and put on a sound mathematical basis by 
Braak \cite{BRA:2013}. Other interesting applications of continued fractions to quantum optics include, for instance, representation of a 
perturbative series for an anharmonic oscillator \cite{CIZ:1984}, solution of master equations in phase space \cite{GAR:2004}, and solution 
of a semiclassical master equation for an atom in a strong electromagnetic field \cite{STE:1972, VAL:1978}. 

Since the study of open quantum systems is a subject with a decades-long history, the continued-fraction approach is only one among many 
methods that came to light over the years. Numerous books and reviews, for example \cite{SPO:1978}--\cite{DVE:2017}, appeared on Markovian 
master equations, non-Markovian master equations of integro-differential, time-convolutionless, and time-discrete type, as well as stochastic 
evolution equations, projection-operator techniques, path-integral procedures, field-theoretic formalisms, and of course numerical endeavours. 
Evidently, the foregoing enumeration is not meant to be exhaustive.  

\section{Kraus map for the density matrix}  

The evolution in time $t$ of a quantum system $S$ that exchanges energy with a thermal reservoir $R$ is completely described by the density 
matrix $\rho_S(t)$, given by   
\begin{equation} 
\rho_{S}(t) = \mbox{Tr}_R \left[ \exp (iH_0 t)\exp (-iH_{SR}t) \rho_{SR} \, \exp (iH_{SR}t) \exp (-iH_0 t) \right] \, , \label{R1}
\end{equation}   
where the interaction picture has been adopted. The initial state of system and reservoir is denoted as $\rho_{SR}$. The Hamiltonian $H_{SR}$ 
governing the unitary dynamics of system and reservoir is equal to $H_0 + H_1$. The free Hamiltonian $H_0$ is equal to the sum of the 
Hamiltonian $H_S$ of the system and the Hamiltonian $H_R$ of the reservoir. The Hamiltonian $H_1$, which describes the interaction between 
system and reservoir, can be expanded as 
\begin{equation} 
H_1 = \sum_{\alpha} V_{\alpha} \otimes U_{\alpha}\, . \label{R2} 
\end{equation} 
In view of assumption (i) of the Introduction, the index $\alpha$ takes on a countable number of values. The system potential $V_{\alpha}$ 
and the reservoir potential $U_{\alpha}$ evolve as 
\[ 
V_{\alpha} (t) = \exp (iH_St) V_{\alpha} \exp (-iH_St) \, ,  
\] 
\begin{equation} 
U_{\alpha} (t) = \exp (iH_Rt) U_{\alpha} \exp (-iH_Rt) \, .  
\label{R3} 
\end{equation} 
Note that the potentials figuring in (\ref{R2}) need not be self-adjoint. In turning (\ref{R1}) into a Kraus map we let us be guided by the 
treatment presented in \cite{EPL:2013}.  

Assumption (ii) of the Introduction allows us to factorize the initial state $\rho_{SR}$ of system and reservoir as $\rho_S \otimes \rho_R$, 
where $\rho_S$ denotes the initial state $\rho_S(t=0)$ of the system. The evolution of the reservoir starts from the thermal state 
$\rho_R= \exp (-\beta H_R)/Z$ of temperature $\beta^{-1}$, with $Z$ equal to $\mbox{Tr}_R [\exp (-\beta H_R)]$. We scale all Hamiltonians 
as well as $\beta^{-1}$ with Planck's constant.  

Upon expanding the unitary evolution operators of (\ref{R1}) as 
\begin{equation} 
\hspace{-18mm}  
\exp (iH_0t) \exp (-iH_{SR}t)  =   1_S \otimes 1_R 
 \label{R4}
\end{equation}  
\[
\hspace{-18mm}  
 + \sum_{m=1}^{\infty} \sum_{\alpha_1 \cdots \alpha_m} (-i )^m \int_0^t \mbox{d} t_1 
\cdots \int_0^{t_{m-1}} \mbox{d} t_m V_{\alpha_1}(t_1) \cdots V_{\alpha_m}(t_m) 
 \otimes  U_{\alpha_1}(t_1) \cdots U_{\alpha_m}(t_m) \, ,  
\]  
we meet reservoir correlation functions of arbitrary order. Owing to assumption (iii) of the Introduction, all of these 
can be factorized by means of Wick's theorem. This gives rise to three types of pair correlation functions, viz.    
\begin{eqnarray} 
c_{\alpha_1\alpha_2} (t_1,t_2) & = & \mbox{Tr}_R\left[ U_{\alpha_1}(t_1) U_{\alpha_2}(t_2)\rho_R\right] \, , \nonumber \\
c_{\alpha_1\alpha_2}^{(+)}(t_1,t_2) & = &  c_{\alpha_1\alpha_2} (t_1,t_2) \, \theta (t_1-t_2) + 
c_{\alpha_2\alpha_1} (t_2,t_1) \, \theta (t_2-t_1) \, , 
\nonumber \\  
c_{\alpha_1\alpha_2}^{(-)}(t_1,t_2) & = & c_{\alpha_1\alpha_2} (t_1,t_2) \, \theta (t_2-t_1) + 
c_{\alpha_2\alpha_1} (t_2,t_1) \, \theta (t_1-t_2) \, ,   
\label{R5} 
\end{eqnarray} 
where $\theta (t)$ denotes the Heaviside step function, i.e., $\theta (t)=1$ for $t>0$ and $\theta (t)=0$ for $t<0$. 

The Wick factorization permits us to generate the dynamics (\ref{R1}) with the help of the non-Markovian map 
\cite{EPL:2013,DIO:2014} 
\begin{equation} 
\rho_S(t) = {\cal T} \exp [ L(t) ]\rho_S \, .  
\label{R6} 
\end{equation}   
The superoperator $L(t)$ comes out as 
\[ 
\hspace{-8mm}
L(t)\rho_S  =  K^{(+)} (t)\rho_S + \rho_S K^{(-)} (t) + \sum_{\alpha\beta} \int_0^t \mbox{d}u \int_0^t \mbox{d}v 
\, c_{\beta\alpha} (v,u) \, V_{\alpha} (u) \rho_S V_{\beta} (v)\, , 
\] 
\begin{equation} 
\hspace{-8mm}  
K^{(\eta)} (t)  =   - {\textstyle \frac{1}{2}} \sum_{\alpha\beta} \int_0^t \mbox{d} u  
\int_0^t \mbox{d} v \, \, c^{(\eta)}_{\alpha\beta}(u,v)\, 
{\cal T}_{\eta} \left \{ V_{\alpha}(u) V_{\beta}(v) \right \} \, . 
\label{R7}
\end{equation} 
The prescription ${\cal T}$ orders products of system potentials $\{V_{\alpha}(t)\}_{\alpha}$ according to 
\begin{eqnarray} 
{\cal T}\left\{ \prod_{i=1}^m V_{\alpha_i}(t_i) \rho_S \prod_{j=1}^n V_{\alpha'_j}(t'_j)\right\} & = &  
{\cal T}_+ \left\{ \rule{0mm}{7mm} \prod_{i=1}^m  V_{\alpha_i}(t_i) \right\} \rho_S 
{\cal T}_- \left\{ \prod_{j=1}^n V_{\alpha'_j}(t'_j) \right\}\,, \nonumber \\  
{\cal T}_+\left\{ \rule{0mm}{7mm} \prod_{i=1}^m V_{\alpha_i}(t_i)\right\}  & = &   
V_{\alpha_1}(t_1)  \cdots V_{\alpha_m}(t_m)\, , \nonumber \\ 
{\cal T}_-\left\{ \rule{0mm}{7mm} \prod_{j=1}^n V_{\alpha'_j}(t'_j)\right\} & = & 
V_{\alpha'_n}(t'_n) \cdots  V_{\alpha'_1}(t'_1)\, , 
\label{R8}
\end{eqnarray} 
where the inequalities $t_1 >  \cdots > t_m$ and $t'_1 >  \cdots  > t'_n$ are assumed.   

By introducing the Kraus matrices  
\begin{equation} 
W_q^{(\eta)} (t;t_1, \cdots,t_q)_{\alpha_1\cdots\alpha_q} = {\cal T}_{\eta} 
\left \{ \exp [ K^{(\eta)} (t)] \prod_{i=1}^q V_{\alpha_i}(t_i)\right \}\, ,   
\label{R9}
\end{equation} 
we can cast the map (\ref{R6}) into the Kraus format \cite{KRA:1971} 
\[
\hspace{-22mm}
\rho_S(t) = W_0^{(+)} (t) \rho_S W_0^{(-)} (t)     
+ \sum_{q=1}^{\infty} \sum_{\alpha^{}_1  \cdots \alpha^{}_q} 
\sum_{\alpha'_1  \cdots \alpha'_q} \int_0^t  \mbox{d} t^{}_1  \cdots \int_0^{t^{}_{q-1}}  \mbox{d} t^{}_q  
\int_0^t \mbox{d} t'_1  \cdots  \int_0^{t'_{q-1}}  \mbox{d} t'_q     
\] 
\[ 
\hspace{-22mm}
\times W_q^{(+)} (t;t^{}_1, \cdots,t^{}_q)_{\alpha^{}_1\cdots\alpha^{}_q} \rho_S 
W_q^{(-)} (t;t'_1,\cdots,t'_q)_{\alpha'_1\cdots\alpha'_q}   
\sum_{P Q} \frac{1}{q!} \prod_{k=1}^q 
c_{\alpha'_{Q(k)}\alpha^{}_{P(k)}} (t'_{Q(k)},t^{}_{P(k)}) \, .
\]  
\vspace{-5mm} 
\begin{equation}  
\label{R10}  
\end{equation} 
On the right-hand side we sum over all permutations $P$ and $Q$ of the integers $\{1,\ldots,q\}$. In order to 
get access to the dissipative dynamics described by (\ref{R10}) one must find a way to cope with the 
time-ordering operator appearing in (\ref{R9}). 

Starting from a Kato identity \cite{KAT:1980}, one shows that the Kraus matrices $\{W_q^{(+)}\}_{q\ge 0}$ satisfy 
the infinite hierarchy 
\[
W_q^{(+)} (t;t^{}_1, \cdots,t^{}_q)_{\alpha^{}_1\cdots\alpha^{}_q} =  V_{\alpha^{}_1}(t^{}_1) 
W_{q-1}^{(+)} (t^{}_1;t^{}_2,\cdots,t^{}_q)_{\alpha^{}_2\cdots\alpha^{}_q}  
\] 
\[ 
-\sum_{j=1}^{q+1}\, \sum_{\alpha \beta}\, \int_{t^{}_1}^t 
\mbox{d} u \int_{t_j}^{t_{j-1}} \mbox{d} v \, c_{\alpha \beta}(u,v) V_{\alpha}(u) 
\] 
\begin{equation}  
\hspace{0mm}
\times W_{q+1}^{(+)} (u;t^{}_1,\cdots,t^{}_{j-1},v,t^{}_j,\cdots,t^{}_q)_{\alpha^{}_1\cdots \alpha^{}_{j-1}\beta 
\alpha^{}_j\cdots\alpha^{}_q}\, , 
\label{R11}  
\end{equation} 
with $t>t_1> \cdots >t_q>0$ and $q\ge 0$. In evaluating the boundaries of the integral over $v$ one has to 
choose $t_0=u$ and $t_{q+1}=0$. If $q$ equals zero on the right-hand side of (\ref{R11}), one must replace 
the contribution $V_{\alpha^{}_1}(t^{}_1) W_{q-1}^{(+)}$ by the unit matrix and the time $t_1$ by zero. 
The Kraus matrices $\{W_q^{(-)}\}_{q\ge 0}$ can be obtained by taking the adjoint of (\ref{R9}) for the case 
$\eta=+$ and replacing the potentials $\{V_{\alpha_i}(t_i)\}_{i=1}^{q}$ by their adjoints. 

Obviously, the effectiveness of the Kraus approach critically depends on the possibility of explicitly computing 
Kraus matrices. In the next section we shall show how the exact solution of the Kraus hierarchy (\ref{R11}) can be 
constructed.  

\section{Exact theory}  

\subsection{Definitions and conventions} 

Before entering into the computation of Kraus matrices, we first propose some definitions and notational conventions. 
The eigenvalues and eigenstates of the system's Hamiltonian $H_S$ are denoted as $\{\omega_k\}_{k\ge 1}$ and 
$\{|k\rangle \}_{k\ge 1}$. The latter provide an orthonormal basis for the Hilbert space of $S$. The representation of 
system potentials and reservoir potentials in terms of the states $\{|k\rangle \}_{k\ge 1}$ reads 
\begin{equation} 
\alpha \rightarrow (kl)\, , \,\,\,\,\,\,\, V_{\alpha}(t) \rightarrow |k\rangle\langle l| \exp [i\omega_{(kl)}t]\, , 
\,\,\,\,\,\,\,  U_{\alpha}(t) \rightarrow U_{(kl)} (t)\, . 
\label{R12}
\end{equation}      
For differences between energy eigenvalues the notation $\omega_{(kl)} = \omega_k - \omega_l$ is used.

Multi-indices of matrices are abbreviated as  
\begin{equation} 
\hspace{-15mm} 
K^n_q = (k_{n+1}k_{n+2}\cdots k_q), \,\,\,\,\,\,\,  K^m_p k K^n_q = (k_{m+1}k_{m+2}\cdots k_pkk_{n+1}k_{n+2}\cdots k_q),  
\label{R13} 
\end{equation} 
with the special case $K_q = K^0_q$. The adjoint of a multi-index matrix $M$ is defined as 
$(M^{\dagger})_{K_qL_q}= M^{\ast}_{L_qK_q}$. For the elements of the multi-index unit matrix the notation   
$\delta_{K_qL_q} = \prod_{s=1}^q \delta_{k_sl_s}$ is employed. 

Time arguments of multivariate functions are abbreviated as 
\begin{equation} 
\hspace{-15mm} 
T^n_q = t_{n+1}, t_{n+2}, \cdots , t_q\, , \,\,\,\,\,\,\, 
T^m_p, t, T^n_q = t_{m+1}, t_{m+2}, \cdots , t_p, t, t_{n+1}, t_{n+2}, \cdots , t_q\, , 
\label{R14} 
\end{equation} 
with the special case $T_q = T^0_q$. If a fixed time $t$ is substracted from all variables of $T^n_q$ we use the 
notation 
\begin{equation} 
T^n_q - t = t_{n+1}-t, t_{n+2}-t, \cdots , t_q-t\, . 
\label{R15} 
\end{equation} 
Repeated integrals over the variables $T^n_q$ are expressed as 
\begin{equation} 
\int_s^t \mbox{d} T^n_q = \int_s^t \mbox{d} t_{n+1} \int_s^{t_{n+1}} \mbox{d} t_{n+2} \cdots \int_s^{t_{q-1}} \mbox{d}t_q\, ,     
\label{R16} 
\end{equation} 
where on the left-hand side the condition $t > t_{n+1} > t_{n+2} > \cdots > t_q > s$ is always in force. 

In the setting (\ref{R12}) the reservoir pair correlation function $c_{\alpha\beta}(t,s)$ takes on the generic form 
$c_{(kl)(mn)}(t,s)$. Laplace transformation of the latter function happens via the standard prescription  
$\hat{c}_{(kl)(mn)}(y) = - i \int_0^{\infty} \mbox{d} t \exp (iyt) c_{(kl)(mn)}(t,0)$, with $\mbox{Im}y$ positive. 
The inverse transform reads $c_{(kl)(mn)}(t,0) = i (2\pi)^{-1} \int_C \mbox{d} y \exp (-iyt) \hat{c}_{(kl)(mn)}(y)$, 
with the contour $C$ determined by $-\infty < \mbox{Re}y < \infty$ and $\mbox{Im}y > 0 $ fixed. 

For the Kraus matrices we switch over to the representation (\ref{R12}) via the definition    
\begin{equation} 
\hspace{-0mm} 
\langle k_1| W_q^{(+)} (t;t^{}_1, \cdots ,t^{}_q)_{(l_1k_2)\cdots (l_qk_{q+1})}|l_{q+1}\rangle = 
W_q(t;T_q)_{K_{q+1}L_{q+1}} \, .
\label{R17} 
\end{equation}   
In (\ref{R9}) the transition $\alpha_j \rightarrow (l_jk_{j+1})$ has been made for $1\le j \le q$. In the new representation 
the Kraus matrices obey the initial condition $W_q(t=0;T_q=0)_{K_{q+1}L_{q+1}} = \delta_{K_{q+1}L_{q+1}}$.   

In order to examine Kraus matrices in Laplace representation we define the following Laplace transform of a multivariate 
matrix $M$ of time arguments $t$ and $T_q$:  
\[
\hspace{-0mm}
\check{M} (z;Z_q)_{K_{q+1}L_{q+1}} =  
\]   
\begin{equation}  
\hspace{-0mm}
(-i)^{q+1} \int_0^{\infty} \mbox{d}t  \int_0^t \mbox{d} T_q 
\exp [izt - i\omega_{k_1}t + iZ_q \cdot T_q] M (t;T_q)_{K_{q+1}L_{q+1}}\, ,  
\label{R18} 
\end{equation}   
where $\mbox{Im}z$ and $\mbox{Im}z_i$ must be positive, with $1\le i\le q$. Since the Laplace variable $z$ is shifted by an 
amount of $- \omega_{k_1}$, the right-hand side of (\ref{R18}) differs from the standard prescription for a Laplace transform. 
The definition of multiple Laplace variables such as $Z_q$ is found from (\ref{R14}) by replacing all uppercase 
letters $T$ by $Z$ and all lowercase letters $t$ by $z$. For the choice $q=0$ the integral over $T_q$ must be omitted. 

The inverse of the transformation (\ref{R18}) contains a multiple complex integral over the single variable $z$ and multiple 
variable $Z_q$. This integral is evaluated by means of repeated integration. By convention, the transform (\ref{R18}) of the 
Kraus matrix $W_{-1}$ is equal to unity, i.e. $\check{W}_{-1} = 1$. By performing in (\ref{R18}) the integrals in reverse order 
and by making use of the initial condition for the Kraus matrices we can evaluate the transformed Kraus matrices for $|z|$ and 
$|z_i|$ large, with $1\le i\le q$. This leads to the asymptotic result 
\begin{equation} 
\check{W}_q (z;Z_q)_{K_{q+1}L_{q+1}} \simeq \prod_{n=0}^q (Z^+_n-\omega_{k_1})^{-1} \delta_{K_{q+1}L_{q+1}}\, , 
\label{R21} 
\end{equation} 
where $\mbox{Im}z$ and $\mbox{Im}z_i$ must be positive. In (\ref{R18}) and (\ref{R21}) the shorthand notation 
\begin{equation} 
Z_q \cdot T_q = \sum_{i=1}^q z_i t_i\, ,  \,\,\,\,\,\,\, Z^+_n = z + z_1 + z_2 + \cdots + z_n\, , 
\label{R19}
\end{equation}   
is employed. 
  
Last, for $q=1,2,3, \dots$ we introduce a superoperator $\Delta_q$, which acts on an arbitrary matrix $\phi (s,s')$ according 
to the rule 
\[  
\Delta_q [\phi ] (t,t';t^{}_1,t'_1;s,s')_{(k^{}_1 l^{}_1)(k'_1 l'_1)} = 
\sum_{K^1_{q+1} L^1_{q+1} K'^1_{q+1} L'^1_{q+1}} \int_s^{t_1} \mbox{d}{T^{}}^1_q \int_{s'}^{t'_1} \mbox{d}{T'}^{1}_q 
\]
\[
\times W^{}_q (t-s ; T_q-s)_{K^{}_{q+1}L^{}_{q+1}} \phi (s,s')_{(l^{}_{q+1})( l'_{q+1})} W^{\dagger}_q (t'-s' ; T'_q-s')_{L'_{q+1}K'_{q+1}}  
\]
\begin{equation}
\times \sum_{PQ} \frac{1}{q!} \prod_{h=1}^q \{Q(h)'\,\,P(h)\}\, .  
\label{5} 
\end{equation} 
We sum over all permutations $P$ and $Q$ of the integers $\{1,2,\dots ,q\}$. The summand contains a product of reservoir pair 
correlation functions. With the help of the notation  
\begin{equation} 
\{m' \,\, n \} = c_{(k'_{m +1}l'_{m})(l^{}_{n}k^{}_{n +1})} (t'_{m}, t^{}_{n})\,   
\label{R20}    
\end{equation} 
this product has been put into a concise form.  

\subsection{Exact Kraus matrices in terms of continued fractions} 

Our program for solving the Kraus hierarchy (\ref{R11}) consists of three parts. In part one we represent the evolution equations 
generating the dissipative dynamics, i.e., the Kraus map (\ref{R10}) and the Kraus hierarchy (\ref{R11}), with the help of the 
eigenstates of the system's Hamiltonian. This representation has the advantage that we dispose of the system potentials (\ref{R3}). 
In part two we make the transfer to Laplace representation so as to remove the time integrals from the Kraus hierarchy. Part one 
and part two of our program are fairly straightforward and lead to a three-term recursion relation for the Kraus matrices. Its 
solution is very lengthy and virtually untractable.  

Therefore, our program is in need of a nontrivial third part. We embed the Kraus matrices in a larger set of matrices, to be 
called matrix ratios. By executing a series of technical maneuvers in the three-term recursion relation for the Kraus matrices, we 
acquire an infinite but closed hierarchy for the matrix ratios. Construction of the iterative solution of the new hierarchy gives 
rise to matrix continued fractions. Explicit derivation of exact expressions for the Kraus matrices is a separate job that is not 
completed in the present treatment.    

With the preparations of subsection 3.1 made, the Kraus map (\ref{R10}) for the evolution of $\rho_S (t)$ can be put into the 
form      
\[
\hspace{-0mm} 
\langle k^{}_1 | \rho_S (t) | k'_1 \rangle = \sum_{l^{}_1 l'_1} W^{}_0 (t)_{(k^{}_1)( l^{}_1)}\, \langle l^{}_1|\rho_S(0)| l'_1 \rangle \,  
W^{\dagger}_0 (t)_{(l'_1)( k'_1)}  
\]
\begin{equation} 
\hspace{24mm} 
+ \sum_{q=1}^{\infty} \sum_{l^{}_1 l'_1} \int_0^t \mbox{d}t^{}_1 
\int_0^t \mbox{d} t'_1 \,\, \Delta_q [\psi ] (t,t;t^{}_1,t'_1;0,0)_{(k^{}_1 l^{}_1)(k'_1 l'_1)}\,\,\, ,  
\label{4} 
\end{equation} 
with the initial condition $\psi(0,0)_{(l^{}_1)(l'_1)}=\langle l^{}_1|\rho_S(0)| l'_1 \rangle $ and $\rho_S(t=0)$ the initial state 
of $S$. Use has been made of the eigenstates $\{\omega_k\}_{k\ge 1}$ of the Hamiltonian $H_S$ and the transition (\ref{R12}) 
for system potentials. In this representation the Kraus hierarchy (\ref{R11}) appears as   
\[
\hspace{-0mm} 
W_q (t;T_q)_{K_{q+1}L_{q+1}} = \delta_{k_1l_1} \exp [i\omega_{(k_1k_2)}t_1] W_{q-1}(t_1;T^1_q)_{K^1_{q+1}L^1_{q+1}}     
\] 
\[
\hspace{-0mm} -\sum_{j=1}^{q+1}\sum_{klm} \int_{t_1}^t \mbox{d} u \int_{t_j}^{t_{j-1}} \mbox{d} v  
\exp [i\omega_{(k_1k)}u]  
\]
\begin{equation} 
\times W_{q+1} (u;T_{j-1},v,T^{j-1}_q)_{(k K^1_j m K^j_{q+1})(L_{j-1} l L^{j-1}_{q+1})} c_{(k_1 k)(lm)}(u,v)\, ,  
\label{8}   
\end{equation}
with $t>t_1>t_2> \cdots >t_q>0$ and $q\ge 0$. In evaluating the boundaries of the integral over $v$ one has to make 
the choices $t_0=u$ and $t_{q+1}=0$. For the choice $q=0$ one must set $t_1$ equal to zero.  

Part one of our program being completed, we turn the hierarchy (\ref{8}) into semi-algebraic form by carrying out the Laplace 
transformation (\ref{R18}). The transformed Kraus hierarchy reads 
\begin{equation}
\hspace{-24mm}
\check{W}_q (z;Z_q)_{K_{q+1}L_{q+1}} = \delta_{k_1 l_1} (z-\omega_{k_1})^{-1} 
\check{W}_{q-1} (z+z_1 ;Z^1_q)_{K^1_{q+1}L^1_{q+1}} 
\label{10}   
\end{equation} 
\[
\hspace{-24mm} 
- \sum_{j=1}^{q+1} \sum_{klm} \int_C \frac{\mbox{d}y}{2\pi i}\, (z-\omega_{k_1})^{-1}   
\check{W}_{q+1} (z-y  ; Z^{}_{j-1}, y, Z^{j-1}_q)_{(kK^{1}_j m K^j_{q+1})(L^{}_{j-1}lL^{j-1}_{q+1})}  
\hat{c}_{(k_{1}k)(lm)}(y) \, .  
\]
The contour $C$ must be parametrized as $-\infty < \mbox{Re}y < \infty$, with $\mbox{Im}y $ fixed and $\mbox{Im}z > \mbox{Im}y >0$. 
Note that $\check{W}_{q-1}$, $\check{W}_q$ and $\check{W}_{q+1}$ appear in the same relation. This three-term recurrence gives rise 
to a bulky solution of little practical value.   

We now embark on the third part of our program. Repeated employment of (\ref{10}) delivers the more general hierarchy 
\[
\hspace{-18mm}
\check{W}_q (z;Z_q)_{K_{q+1}L_{q+1}} = \delta_{K_n L_n} \prod_{s=1}^n (Z^+_{s-1}-\omega_{k_s})^{-1} 
\check{W}_{q-n} (Z^+_n ;Z^n_q)_{K^n_{q+1}L^n_{q+1}}  
\] 
\[
\hspace{-18mm} 
-\sum_{p=-1}^{n-2} \sum_{j=p+2}^{q+1} \sum_{klm} \int_C \frac{\mbox{d}y}{2\pi i}\,  \delta_{K_{p+1}L_{p+1}} 
\prod_{s=1}^{p+2} (Z^+_{s-1}-\omega_{k_s})^{-1}   
\]
\begin{equation}  
\hspace{-18mm}
\times \check{W}_{q-p} (Z^+_{p+1} -y  ; Z^{p+1}_{j-1}, y, Z^{j-1}_q)_{(kK^{p+2}_j m K^j_{q+1})(L^{p+1}_{j-1}lL^{j-1}_{q+1})}  
\hat{c}_{(k_{p+2}k)(lm)}(y) \, , 
\label{11} 
\end{equation}  
where the conditions $q\ge 0$ and $0\le n \le q+1$ must be satisfied. As it turns out, the hierarchy (\ref{10}) does not allow 
for further treatment on the basis of continued fractions, whereas the extended hierarchy (\ref{11}) does. 

To see this in detail we introduce the following matrix ratio:  
\[  
\hspace{-0mm}
R_{q,n} (z;Z_q)_{K_{q+1}L_{q+1}} =  
\] 
\begin{equation}  
\hspace{-0mm}
\sum_{M_{q+1}} \check{W}_q (z; Z_q)_{K_{q+1}M_{q+1}} \delta_{M_nL_n} \check{W}_{q-n}^{-1} 
(Z^+_n ; Z^n_q)_{M^n_{q+1}L^n_{q+1}} \, , 
\label{13} 
\end{equation} 
with $q \ge 0$ and $0\le n\le q+1$. The imaginary parts of the Laplace variables $z$ and $\{z_i\}_{i=1}^q$ must be chosen 
sufficiently large. Then the determinant $|\check{W}_{q-n} (Z^+_n ; Z^n_q)|$ differs from zero, as 
follows from (\ref{R21}) and a continuity argument. Therefore, the right-hand side of (\ref{13}) is well-defined in the entire 
domain that is needed for carrying out inverse Laplace transformation. Note that for $n=0$ the matrix ratio 
reduces to $R_{q,0}(z;Z_q)_{K_{q+1}L_{q+1}} = \delta_{K_{q+1}L_{q+1}}$.  The equality 
\begin{equation} 
R_{q,q+1} (z;Z_q)_{K_{q+1}L_{q+1}} = \check{W}_q (z; Z_q)_{K_{q+1}L_{q+1}}
\label{14}
\end{equation} 
is a consequence of the convention $\check{W}_{-1} = 1$. It shows that the set of Kraus matrices is completely contained in the 
larger family of matrix ratios. 

Upon multiplying (\ref{11}) from the right by $\check{W}_q^{-1}$, the unit matrix appears on the left-hand side, whereas in the 
first contribution on the right-hand side the inverse of the matrix ratio (\ref{13}) is created. The second term on the 
right-hand side can be restated in terms of matrix ratios if we insert the identity 
$\sum_{M_{q+1}} G^{-1}(z;Z_q)_{K_{q+1}M_{q+1}}G(z;Z_q)_{M_{q+1}L_{q+1}}= \delta_{K_{q+1}L_{q+1}}$, with the intermediate matrix 
$G$ chosen as  
\begin{equation}  
G(z;Z_q)_{K_{q+1}L_{q+1}} = \delta_{K_jL_j} \check{W}_{q-j} (Z^+_j ; Z^j_q)_{K^j_{q+1}L^j_{q+1}}\, . 
\label{15} 
\end{equation}  
Rearrangement of dummy variables brings us to the key result 
\[
\hspace{-24mm}
R^{-1}_{q,n} (z; Z_q)_{K_{q+1}L_{q+1}} =  \delta_{K_{q+1}L_{q+1}} \prod_{s=1}^n (Z^+_{s-1} - \omega_{k_s}) 
+ \sum_{p=-1}^{n-2} \sum_{j={p+2}}^{q+1} \sum_{M_{q+1}} \sum_{klm} \int_C \frac{\mbox{d}y}{2\pi i}  \delta_{K_{p+1}M_{p+1}}    
\] 
\[  
\hspace{-24mm}
\times  \prod_{s=p+3}^n (Z^+_{s-1} - \omega_{k_s}) 
R_{q-p, j-p} (Z^+_{p+1}-y; Z^{p+1}_{j-1},y,Z^{j-1}_q)_{(kK^{p+2}_jmK^j_{q+1})(M^{p+1}_{j-1}lM^{j-1}_{q+1})} 
\]
\begin{equation}  
\hspace{-24mm}
\times  R^{-1}_{q,j}(z;Z_q)_{M_{q+1}L_{q+1}} \hat{c}_{(k_{p+2}k)(lm)}(y)   \, ,
\label{16}
\end{equation}   
with the constraint $0\le n \le q+1$. For $n=0$ the right-hand side of (\ref{16}) reduces to the unit matrix, in line with 
definition (\ref{13}). 

Matrix inversion on both sides of (\ref{16}), followed by iteration and the choice $n=q+1$, furnishes the exact Kraus matrices. 
That is to say, the Kraus matrices emanating from the Schr\"{o}dinger equation for system and reservoir in the setting (i)-(iii) 
of the Introduction. As planned, our solution is built up from matrix continued fractions. It constitutes a direct extension of 
the standard formula (\ref{1}) to the case of a multi-level system interacting with a reservoir of finite temperature. However, 
execution of the afore-mentioned program requires a lot of technical effort and is therefore shifted to another paper 
\cite{arXiv:2017}. In the present treatment, we shall focus on the derivation of an approximate continued-fraction solution 
possessing a much simpler structure and allowing for analytical work. This is the subject of the next section. 

Before closing we solve the Kraus hierarchy for a damped two-level atom at zero temperature. We couple the atom to the 
annihilation and creation operators $b(\omega)$ and $b^{\dagger}(\omega)$ of the radiation field via the interaction Hamiltonian 
$|1\rangle \langle 2| \otimes \int_0^{\infty} \mbox{d}\omega g^{\ast}(\omega ) b^{\dagger} (\omega) + \mbox{h.c.}$, where 
counter-rotating contributions have been dropped, as well as the diagonal reservoir potentials $U_{(11)}$ and $U_{(22)}$. Then 
all correlation functions $c_{(kl)(mn)}$ equal zero, except for the choice $k=n=2,l=m=1$. Furthermore, the atom is assumed 
to start from the excited state so that the initial condition $\rho_S = |2\rangle \langle 2|$ is in force. Now the matrix 
$\check{W}_{q+1}$ figuring on the right-hand side of the Kraus hierarchy (\ref{10}), with 
$K_{q+1}= (12\cdots 22)$, $L_{q+1}= (11\cdots 12)$ and $q\ge 0$, can be expressed in terms of the matrix $\check{W}_q$, with 
$K_q=(2\cdots 22)$ and $L_q=(1\cdots 12)$. It then appears that $\check{W}_q$ is vanishing for $q\ge 1$. Hence, in (\ref{4}) 
only the term with $W_0$ survives. From the Kraus hierarchy (\ref{10}) we obtain the solution    
\begin{equation} 
\check{W}_0(z)_{(2)(2)} = 
\left[ z-\omega_2 + \int_C \frac{\mbox{d}y}{2\pi i} \frac{\hat{c}_{(21)(12)}(y)}{ z - y -\omega_1 } \right]^{-1} \, .   
\label{16A} 
\end{equation}  
The contour $C$ must satisfy the condition $\mbox{Im} z > \mbox{Im} y > 0$.  At zero temperature 
the Laplace transformed correlation function is found as 
\begin{equation} 
\hat{c}_{(21)(12)}(y) = \int_0^{\infty} \mbox{d} \omega |g (\omega)|^2 / (y-\omega )\, .  
\label{16B} 
\end{equation} 
Upon elaborating the first contribution of (\ref{4}) with the help of (\ref{16A}) and (\ref{16B}) we recover the well-known result 
(\ref{1}). 

\section{Perturbation theory} 

In developing a sound perturbation theory, we should safeguard the properties $\rho_S(t)\ge 0$ and $\mbox{Tr}\rho_S(t)=1$ 
of the exact density matrix. Only then it is guaranteed that the perturbative dynamics does not exhibit unphysical artefacts. 
Furthermore, we should work in arbitrary perturbative order $N$, the limit $N\rightarrow \infty$ being capable of generating 
the exact dynamics. This permits us to control errors coming along with the perturbative approach. Realization of our 
perturbative program will be achieved in three steps: (i) truncation of the Kraus hierarchy; (ii) factorization of Kraus matrices 
of order $N$ and higher; (iii) factorization of the permutations $P$ and $Q$ figuring in (\ref{5}).  

First, we truncate the Kraus hierarchy through discarding the second term on the right-hand side of (\ref{8}) for $q=N$ and 
replacing $\delta_{k_1l_1}$ by $W_0(t-t_1)_{(k_1)(l_1)}$. Thus the truncation prescription is given by the factorization  
\begin{equation} 
\hspace{-16mm} 
W_N (t;T_N)_{K_{N+1}L_{N+1}} \rightarrow  \exp [i\omega_{(k_1k_2)}t_1] W'_0(t-t_1)_{(k_1)(l_1)}  
W'_{N-1}(t_1;T^1_N)_{K^1_{N+1}L^1_{N+1}}\, ,  
\label{17}
\end{equation}
where the ordering $t>t_1>t_2>\cdots >t_N>0$ is in force and a prime is used to denote perturbative Kraus matrices 
\footnote{In \cite{EPL:2013} a tilde was used to distinguish between exact and perturbative Kraus matrices. This notation becomes 
awkward for the Laplace transform $\check{W}_0$.}. The truncation parameter $N$ defines the order of the perturbation theory and 
takes on the values $N=1,2,3, \ldots$. In \cite{EPL:2013} it was shown that (\ref{17}) is exact in the asymptotic regime of large 
times.  

Second, for Kraus matrices of higher order we perform the factorization 
\[ 
\hspace{-18mm} 
W_{q+pN}(t;T_{q+pN})_{K_{q+pN+1}L_{q+pN+1}} \rightarrow 
\exp (i\omega_{(k_1k_{q+2})}t_{q+1}) W'_{q}(t-t_{q+1};T_{q}-t_{q+1})_{K_{q+1}L_{q+1}}
\]
\[  
\hspace{-18mm} 
\times \prod_{j=1}^p \left [ \exp (i\omega_{(k_{q+(j-1)N+2}k_{q+jN+2})}t_{q+jN+1}) \right.  
\]  
\begin{equation} 
\hspace{-18mm}   
\left. \times 
W'_{N-1}(t_{q+(j-1)N+1}-t_{q+jN+1};T^{q+(j-1)N+1}_{q+jN}-t_{q+jN+1})_{K^{q+(j-1)N+1}_{q+jN+1}L^{q+(j-1)N+1}_{q+jN+1}} \right ] \, , 
\label{18} 
\end{equation}  
with $q=0,1,2,\ldots,N-1$ and $p=1,2,3,\ldots$. We make use of the convention $t_{q+pN+1}=0$. The right-hand side of (\ref{18}) 
is completely determined by the Kraus matrices $\{W'_q\}_{q=0}^{N-1}$. The latter can be obtained by solving the truncated 
Kraus hierarchy. This closed set of $N$ equations corresponds to the choices $q=0,1,2,\ldots,N-1$ in (\ref{8}). Note that 
(\ref{18}) reduces to (\ref{17}) under the choices $q=0$ and $p=1$, so the first step is consistent with the second step. 
Last, we point out that factorization of the adjoint Kraus matrices happens by taking the complex conjugate of (\ref{18}). 

Third, having fixed all perturbative Kraus matrices, we commence the construction of a perturbative series for the density matrix.  
To that end, (\ref{18}) is substituted into (\ref{4}) and the sum over $q$ is divided into bunches of $N$ terms. Next, $P$ and $Q$ 
are factorized such that any coupling between different sectors of the perturbative series is eliminated. Recalling Wick's theorem, 
we see that this transition comes down to discarding reservoir correlation functions with more than $2N$ time arguments. In other 
words, memory effects induced by the reservoir are taken into account up to a certain degree only.  

Implementation in (\ref{4}) of the factorizations as described above provides us with the perturbative density 
matrix $\rho_S^{(N)}(t)$ we are after. The expansion for $\rho_S^{(N)}(t)$ can be generated with the help of the identity    
\[  
\hspace{-0mm} 
\langle k^{}_1 | \rho^{(N)}_S (t) | k'_1 \rangle = 
\sum_{l^{}_1 l'_1} W'^{}_0 (t)_{(k^{}_1)( l^{}_1)}\, \langle l^{}_1|\rho_S(0)| l'_1 \rangle \,  
W'_0{}^{\, \dagger} (t)_{(l'_1)( k'_1)}  
\]
\[
\hspace{27mm} 
+ \sum_{q=1}^{N-1} \sum_{l^{}_1 l'_1} \int_0^t 
\mbox{d}t^{}_1 \int_0^t \mbox{d} t'_1 \,\, \Delta'_q [\psi ] (t,t ;t^{}_1,t'_1;0,0)_{(k^{}_1 l^{}_1)(k'_1 l'_1)}   
\] 
\[
\hspace{27mm} 
+ \sum_{l^{}_1 l'_1} \int_0^t \mbox{d}s \int_0^t \mbox{d} s' \exp [i\omega_{k^{}_1}s - i\omega_{k'_1}s']    
\]
\[ 
\hspace{27mm} 
\times W'^{}_0 (t-s)_{(k^{}_1)( l^{}_1)}\, \chi^{(N)} (s,s')_{(l^{}_1)( l'_1)}\, W'_0{}^{\, \dagger} (t-s')_{(l'_1)( k'_1)}  
\]
\[
\hspace{27mm} 
+ \sum_{q=1}^{N-1} \sum_{l^{}_1 l'_1} \int_0^t \mbox{d}s \int_0^t \mbox{d} s' 
\int_s^t \mbox{d} t^{}_1 \int_{s'}^t \mbox{d} t'_1 \exp [i\omega_{k^{}_1}s - i\omega_{k'_1}s']  
\] 
\begin{equation}  
\hspace{27mm} 
\times \Delta'_q [\chi^{(N)} ] (t,t ;t^{}_1,t'_1;s,s')_{(k^{}_1 l^{}_1)(k'_1 l'_1)}  \, .   
\label{19}
\end{equation}  
The bitemporal matrix $\chi^{(N)}$ satisfies the integral equation 
\[
\hspace{-8mm} 
\chi^{(N)} (t^{}_1,t'_1)_{(l^{}_1)(l'_1)} = \sum_{k^{}_1 k'_1} 
\exp \left[ -i\omega_{k^{}_1}t^{}_1 + i \omega_{k'_1}t'_1 \right]  
\Delta'_N [\psi ] (t^{}_1,t'_1 ;t^{}_1,t'_1;0,0)_{(k^{}_1 l^{}_1)(k'_1 l'_1)}   
\]
\[
\hspace{22mm} 
+ \sum_{k^{}_1 k'_1} \int_0^{t^{}_1} \mbox{d}s \int_0^{t'_{1}} \mbox{d} s' 
\exp \left[ -i\omega_{k^{}_1}(t^{}_1 -s) + i \omega_{k'_1} (t'_1-s') \right]  
\] 
\begin{equation} 
\hspace{22mm} 
\times \Delta'_N [\chi^{(N)} ] (t^{}_1,t'_1 ;t^{}_1,t'_1;s,s')_{(k^{}_1 l^{}_1)(k'_1 l'_1)} \, ,   
\label{20}
\end{equation} 
with $N=1,2,3, \ldots$ and the initial condition 
$\psi (0,0)_{(l^{}_1)(l'_1)}=\langle l^{}_1|\rho_S(0)| l'_1 \rangle $, as in (\ref{4}). 
Last, the superoperators $\Delta'_q$ and $\Delta'_N$ must be obtained from (\ref{5}) through replacement 
of all Kraus matrices by their primed counterparts.  

The set (\ref{19})--(\ref{20}) determines the evolution of the density matrix of the open quantum system 
$S$ in arbitrary perturbative order $N$. For the case of $N=1$ we carry out the additional transformation 
\begin{equation}
\hspace{-20mm}
\xi (t,t')_{(k)(k')} = \sum_{l l'} W'_0 (t)_{(k)(l)}  \langle l|\rho_S(0)| l' \rangle
W'_0{}^{\, \dagger} (t')_{(l')(k')}  
\label{21}   
\end{equation} 
\[
\hspace{-20mm} 
+ \sum_{l l'} \int_0^t \mbox{d} s \int_0^{t'} \mbox{d} s' 
\exp [i\omega_k s - i\omega_{k'} s'] W'_0(t-s)_{(k)(l)} \chi^{(1)} (s,s')_{(l)(l')}  
W'_0{}^{\, \dagger} (t'-s')_{(l')(k')} \, .   
\]  
Then the representation of the lowest-order perturbative density matrix simplifies to   
$\langle k | \rho^{(1)}_S (t) | k' \rangle = \xi (t,t)_{(k)(k')}$.  The shifted 
bitemporal matrix $\xi$ satisfies the evolution equation   
\[ 
\hspace{-10mm}
\xi (t,t')_{(k)(k')} = \sum_{l l'} W'_0 (t)_{(k)(l)}  \langle l|\rho_S(0)| l' \rangle 
W'_0{}^{\, \dagger} (t')_{(l')(k')} 
\]  
\[
\hspace{16mm} 
+ \sum_{l l' m m'} \int_0^t \mbox{d} s \int_0^{t'} \mbox{d} s' 
\exp [i\omega_{(km)}s + i\omega_{(m'k')}s'] W'_0(t-s)_{(k)(l)}  
\]  
\begin{equation} 
\hspace{16mm} 
\times \xi (s,s')_{(m)(m')}  
W'_0{}^{\, \dagger} (t'-s')_{(l')(k')} \, c_{(m'l')(lm)}(s',s) \, , 
\label{22}  
\end{equation}  
whereas the Kraus hierarchy reduces to the single equation  
\[  
\hspace{6mm} 
W'_0(t)_{(k)(l)} = \delta_{kl} - \sum_{mnp} \int_0^t \mbox{d} u \int_0^u \mbox{d} v 
\exp [i\omega_{(km)} u + i\omega_{(mp)} v] 
\]
\begin{equation}  
\hspace{26mm}
\times W'_0(u-v)_{(m)(n)} W'_0(v)_{(p)(l)} c_{(km)(np)}(u,v) \, .  
\label{23}  
\end{equation}  
The dissipative dynamics governed by (\ref{22}) and (\ref{23}) was first derived in \cite{JPA:2006}.  

After employment of the Laplace transform (\ref{R18}), the matrix $\check{W}'_0 (z)$ appears both on the left-hand 
side and on the right-hand side of (\ref{23}). Therefore, we are invited to multiply this equation from the right 
by $\check{W}'_0{}^{\, -1} (z)$. In view of (\ref{R21}) the afore-mentioned inverse matrix exists for $\mbox{Im}z$ 
sufficiently large. It is found to obey   
\begin{equation} 
\hspace{-0mm} 
\check{W}'_0{}^{\, -1} (z)_{(k)(l)} = 
 (z-\omega_k)\delta_{kl} + \sum_{mn} \int_C \frac{\mbox{d}y}{2\pi i} \hat{c}_{(km)(nl)}(y) 
\check{W}'_0 (z-y)_{(m)(n)}\, ,  
\label{24} 
\end{equation}  
with $\mbox{Im}z > \mbox{Im}y > 0$. This is a finite-temperature identity which is valid in the presence of 
counter-rotating contributions in the interaction Hamiltonian for system and reservoir. Solution of (\ref{24}) 
happens by matrix inversion on both sides and subsequent iteration. The ensuing matrix continued fraction possesses 
a much more orderly structure than its exact counterpart originating from (\ref{16}). At the same time, (\ref{24}) 
is not an ad hoc result  since it is the low-end product of a full-fledged perturbation theory that is tied up 
with the exact dynamics through the limit of $N\rightarrow \infty$.  

The analytic properties of $\check{W}'_0{}(z)$ can be explored by temporarily assuming a discrete reservoir  
with energy eigenvalues $\{\mu_j\}_j$ and energy eigenstates $\{|r_j\rangle \}_j$. Thus the expansion 
$H_R = \sum_j \mu_j |r_j\rangle \langle r_j|$ can be utilized. It then appears that $\check{W}'_0{}(z)$ is analytic 
for $\mbox{Im}z \ne 0$ and that this matrix possesses simple poles $\{z=x_s\}_s$ on the real axis. The foregoing 
statement can be proved by applying induction to the iterative solution for $\check{W}'_0{}(z)$. In doing so, 
(\ref{24}) should be replaced by the matrix identity 
\begin{equation} 
\hspace{-16mm}
\check{W}'_0{}^{\, -1} (z) = E(z) - \sum_{jj's} (z-x_s + \mu_j - \mu_{j'})^{-1} C^{jj'}\cdot 
\left( \frac{\!\!\mbox{d}}{\mbox{d}x_s} \check{W}'_0{}^{\, -1} (x_s) \right )^{-1}\!\! \cdot\, C^{jj'\,\dagger} \, ,     
\label{25} 
\end{equation} 
where the dot means matrix multiplication. The matrices $E(z)_{kl}$ and $C^{jj'}_{kl}$ are defined as 
$(z-\omega_k)\delta_{kl}$ and $\langle r_j|\rho_R |r_j\rangle^{1/2} \langle r_j|U_{(kl)} |r_{j'}\rangle$, respectively. 

While carrying out the above-mentioned induction proof, one verifies in each iterative order that the inverse matrix 
on the right-hand side of (\ref{25}) is strictly positive. Hence, if  $\mbox{Im}z$ differs from zero the imaginary 
part of the standard quadratic form $<v,\check{W}'_0{}^{\, -1} (z) v>$ differs from zero as well, for arbitrary 
vector $v\ne 0$. Consequently, all poles of the matrix $\check{W}'_0{}(z)$ lie on the real axis and inverse Laplace 
transformation can be performed on the basis of the contour $z=\omega +i0$, with $\omega$ real. As one takes a 
continuum limit for the reservoir, this important conclusion remains valid, albeit that the poles on the real axis 
merge together into a branch cut.     
 
Returning to the case of arbitrary perturbative order, we are now going to verify that both positivity and probability 
are conserved for the perturbative density matrix. By construction, the expansion for $\rho_S^{(N)}(t)$ has the Kraus 
form  $\sum_j V_j \rho_S V_j^{\dagger}$, where the matrices $\{ V_j\}_j$ need not be specified. As a consequence, the 
property $\rho_S^{(N)}(t) \ge 0$ is obvious. In contrast, conservation of probability is not manifest for  
$\rho_S^{(N)}(t)$ and indeed the proof requires some effort. In Appendix A the result 
\begin{equation} 
\frac{\mbox{d}}{\mbox{d}t}  \sum_{k_1} \langle k^{}_1 | \rho^{(N)}_S (t) | k_1 \rangle = 0  
\label{26} 
\end{equation}  
is established, with $N=1,2,3, \ldots$.  

Let us again focus on the lowest perturbative order $N=1$ and reconsider the damped two-level atom at zero temperature. 
Choosing the interaction Hamiltonian as specified in subsection 3.2, one recognizes that the solution of (\ref{24}) 
for $k=l=2$ coincides with (\ref{16A}). Hence, the lowest-order perturbation theory is capable of reproducing the exact 
evolution (\ref{1}). Moreover, for a system of arbitrary dimension this theory is also capable of reproducing the 
Markovian map $\rho_S (t) = \mbox{exp} (Lt)\rho_S (0)$. To that end, the evolution equations (\ref{22}) and (\ref{23}) 
must be solved in the limit of large time and weak coupling between system and reservoir. As demonstrated in 
\cite{JPA:2006}, for the constant generator $L$ the standard \cite{DAV:1973} form originating from the exact dynamics 
(\ref{R10}) is obtained. 

For the damped two-level atom of subsection 3.2 the above-mentioned Markovian map provides us with the Wigner-Weisskopf 
evolution 
\begin{eqnarray}
\langle 2 |\rho_S(t)| 2 \rangle  =  1 - \langle 1 |\rho_S(t)| 1 \rangle    
& = &  \exp (-2\gamma t) \langle 2 |\rho_S(0)| 2 \rangle  \, ,
\nonumber \\  
\langle 2 |\rho_S(t)| 1 \rangle  = \,\,\,\, \langle 1 |\rho_S(t)| 2 \rangle^{\ast} 
& = & \exp (-\gamma t + i \bar{\omega} t) \langle 2 |\rho_S(0)| 1 \rangle \, ,  
\label{26A} 
\end{eqnarray}
where we have defined the damping constant $\gamma = \pi |g(\omega_{(21)})|^2$ and the level shift 
$\bar{\omega} = {\cal P}\int_0^{\infty} \mbox{d}\omega |g(\omega)|^2/(\omega - \omega_{(21)})$. In the field of 
quantum communication the evolution (\ref{26A}) is cast into the form of an amplitude-damping channel \cite{IMR:2013}, 
given by   
\begin{equation} 
\rho_S (t) = M(t) \rho_S (0) M(t)^{\dagger} +  N(t) \rho_S (0) N(t)^{\dagger}\, .  
\label{26B} 
\end{equation}   
The nonzero elements of the Kraus matrices read $\langle 1| M(t) |1\rangle =1$, 
$\langle 2| M(t) |2\rangle = \exp (-\gamma t + i\bar{\omega} t)$, and $\langle 1| N(t) |2\rangle = [1-\exp(-2\gamma t)]^{1/2}$. 
 
In the next section, we shall discuss an application of the dissipative quantum theory resting on the evolution equations 
(\ref{22}) and (\ref{23}). We shall demonstrate that density matrices can be computed with relative ease and that 
quantum evolutions can be investigated by analytical means. 
  
\section{Jaynes-Cummings model with non-Markovian damping}  

As system $S$ we choose a two-level atom with excited state $|2\rangle_{a}$ of energy $\omega_{a,2}$ and ground state 
$|1\rangle_{a}$ of energy $\omega_{a,1}$, as well as a single electromagnetic mode with states 
$|n\rangle_{f} = (n!)^{-1/2} a^{\dagger \, n}|0\rangle_{f}$ of energies $n\omega_{f}$ $(n=0,1,2,\ldots)$. Here 
$|0\rangle_{f}$ denotes the vacuum state of the field mode and $a^{\dagger}$ the photon creation operator. The atom is 
assumed to be on resonance with the field mode, so the statement $\omega_{f} = \omega_{a,2} - \omega_{a,1} > 0$ holds 
true. Exchange of energy between atom and field mode happens through the Jaynes-Cummings interaction 
$f |1\rangle_{a} \left._{a}\langle 2|\right. \otimes a^{\dagger} + \mbox{h.c.}$, with $f$ real and positive.  

For the orthonormal energy eigenstates and energy eigenvalues of $S$ we find 
\begin{eqnarray}  
|\epsilon , n\rangle & = & \nu_n ( |1\rangle_a \otimes |n+1\rangle_f + 
\epsilon |2\rangle_a \otimes |n\rangle_f ) \,\,\, ,  \nonumber  \\ 
\Omega_{\epsilon , n} & = & \omega_{a,2} (n+1) -\omega_{a,1}n + \epsilon f (n+1)^{1/2}  
\,\,\, , 
\label{28}
\end{eqnarray}   
with $\epsilon = -1, +1$, $n=-1,0,1,\ldots$, $\nu_n = \delta_{n,-1} + \theta_n/ \sqrt{2}$, and $|-1\rangle_{f} \equiv 0$. One has 
$|-1,-1\rangle = |+1,-1\rangle$, so for $n=-1$ we drop the choice of $\epsilon = -1$. In summations a prime appears to 
indicate this. By $\theta_n$ the discrete theta function is meant, i.e., $\theta_n = 1$ for $n$ nonnegative and $\theta_n = 0$ for 
$n$ negative. The eigenstates $\{ |\epsilon , n \rangle  \}$ satisfy a completeness relation, so they span the system's Hilbert 
space. We assume that the inequality $\omega_f > f$ is true. Then one has the property $\Omega_{\epsilon,n} > \Omega_{+1,-1}$ for 
arbitrary $(\epsilon , n)\ne (+1,-1)$, which expresses the fact that the energy of the state $|+1,-1\rangle $ is lowest.

The above model describes an immobile atom in a cavity that selects a privileged field mode. Radiative damping is introduced by 
coupling the atom to a continuum of transverse electromagnetic modes. These are (de-)excited by the annihilation and creation 
operators $b(\omega)$ and $b^{\dagger}(\omega)$. The transverse electromagnetic continuum is kept at zero temperature. Furthermore, 
cavity damping and collisional damping are not taken into consideration. Discarding counter-rotating contributions, we can model  
the interaction between system and reservoir as  
\begin{eqnarray}
H_1 & = & |1\rangle_{a} \left._{a}\langle 2|\right. \otimes \int_0^{\infty} \mbox{d} \omega \, g^{\ast} (\omega) b^{\dagger}(\omega )  
          + \mbox{h.c.} \, , \nonumber \\    
    & = &  \sum_{\epsilon_1,n_1,\epsilon_2,n_2} \rule{0mm}{5mm}^{\!\!\!\!\!\!\prime}
\,\, |\epsilon_1,n_1\rangle\langle\epsilon_2,n_2| 
\otimes U_{(\epsilon_1,n_1)(\epsilon_2,n_2)}\, , \nonumber \\ 
U_{(\epsilon_1,n_1)(\epsilon_2,n_2)} & = & 
{\textstyle \frac{1}{\sqrt{2}}} \epsilon_2 \nu_{n_1}\theta_{n_2} 
\delta_{n_1+1,n_2} \int_0^{\infty} \mbox{d}\omega \, g^{\ast}(\omega) b^{\dagger}(\omega) \nonumber \\ 
    &   & + {\textstyle \frac{1}{\sqrt{2}}} \epsilon_1 \nu_{n_2}\theta_{n_1} 
\delta_{n_1,n_2+1} \int_0^{\infty} \mbox{d}\omega \, g(\omega) b(\omega) \,\,\, .  
\label{29}  
\end{eqnarray} 
As announced, a prime is used to exclude the state $|-1,-1\rangle$ from the summation.  

We factorize the initial state of $S$ and choose the privileged mode to be in a number state of $p$ photons. Then the matrix 
elements of $\rho_S(0)$ read 
\begin{eqnarray} 
\langle\epsilon_1,n_1|\rho_S(0)|\epsilon_2,n_2\rangle & = & \nu_{n_1}\nu_{n_2} (\rho_{a,11}
\delta_{n_1,n_2}\delta_{n_1+1,p} +  
\epsilon_1\rho_{a,21}\delta_{n_1,n_2+1}\delta_{n_1,p} \nonumber \\ 
 & & + \epsilon_2\rho_{a,12}\delta_{n_1+1,n_2}\delta_{n_2,p} 
 + \epsilon_1\epsilon_2 \rho_{a,22}\delta_{n_1,n_2}\delta_{n_1,p}  )\,\,\, ,   
\label{30} 
\end{eqnarray}  
with $\rho_{a,kl} = \left._{a}\langle k | \rho_a(0) | l \rangle_{a}\right.$ and $\rho_a(0)$ the initial atomic state. The 
pair correlation functions come out as  
\begin{eqnarray}  
c_{((\epsilon_1,n_1)(\epsilon_2,n_2))((\epsilon_3,n_3)(\epsilon_4,n_4))} (t,0) & = &  
{\textstyle \frac{1}{2}} \epsilon_1 \epsilon_4 \theta_{n_1} \theta_{n_4}  
\nu_{n_2} \nu_{n_3} \delta_{n_1,n_2+1} \delta_{n_3+1,n_4} \nonumber \\ 
 &  & \times \int_0^{\infty} \mbox{d} \omega |g(\omega)|^2 \exp (-i\omega t) \,\,\, .
\label{31} 
\end{eqnarray} 
We are ready now to construct the solution of (\ref{22}). 

Making use of the Laplace transform 
\begin{equation}  
\hspace{-0mm} 
\check{\xi} (z,z')_{(\epsilon^{}_1,n^{}_1)(\epsilon'_1,n'_1)} =  
\label{32}
\end{equation}
\[
\hspace{-0mm} 
 \int_0^{\infty} \mbox{d}t \int_0^{\infty} \mbox{d}t' 
\exp (izt-i\Omega_{\epsilon^{}_1,n^{}_1}t - iz't'+i\Omega_{\epsilon'_1,n'_1}t')  
\xi (t,t')_{(\epsilon^{}_1,n^{}_1)(\epsilon'_1,n'_1)} \,\,\, , 
\]  
with $\mbox{Im}z$ and $-\mbox{Im} z'$ positive, we can cast (\ref{22}) into an algebraic form, given by     
\[ 
\hspace{-20mm}
\check{\xi} (z,z')_{(\epsilon^{}_1,n^{}_1+1)(\epsilon'_1,n'_1+1)} =  
\]
\[ 
\hspace{-20mm}
\sum_{\epsilon^{}_2,n^{}_2,\epsilon'_2,n'_2} \rule{0mm}{5mm}^{\!\!\!\!\!\!\prime} 
\check{W}'_0 (z)_{(\epsilon^{}_1,n^{}_1+1)(\epsilon^{}_2,n^{}_2)}  
\langle\epsilon^{}_2,n^{}_2|\rho_S(0)|\epsilon'_2,n'_2\rangle  
\bar{W}'_0 (z')_{(\epsilon'_1,n'_1+1)(\epsilon'_2,n'_2)}  
\] 
\[ 
\hspace{-20mm}
+ \sum_{\epsilon^{}_2,n^{}_2,\epsilon'_2,n'_2, \epsilon^{}_3, \epsilon'_3} 
\rule{0mm}{5mm}^{\!\!\!\!\!\!\!\!\!\!\!\!\!\prime}  
\,\,\,\,\,\,\,\, {\textstyle \frac{1}{2}} \nu_{n^{}_2}\nu_{n'_2} 
\epsilon^{}_3 \epsilon'_3 \int_0^{\infty} \mbox{d}\omega_1 |g(\omega_1)|^2 \, 
\check{W}'_0 (z)_{(\epsilon^{}_1,n^{}_1+1)(\epsilon^{}_2,n^{}_2)}  
\bar{W}'_0 (z')_{(\epsilon'_1,n'_1+1)(\epsilon'_2,n'_2)}  
\]
\begin{equation} 
\hspace{-20mm} 
\times \, \check{\xi} (z +\omega_1\, ,z' + \omega_1)_{(\epsilon^{}_3,n^{}_2+1)(\epsilon'_3,n'_2+1)}  \, ,
\label{33} 
\end{equation}  
with $\mbox{Im}z$ and $-\mbox{Im} z'$ positive. For the transform $\check{W}'_0$ we find from 
(\ref{24}) and (\ref{31}) the relation    
\begin{equation}
\hspace{-0mm}
\check{W}'_0{}^{\, -1} (z)_{(\epsilon_1,n_1)(\epsilon_2,n_2)} = 
(z-\Omega_{\epsilon_1,n_1}) \delta_{\epsilon_1,\epsilon_2}\delta_{n_1,n_2} 
\label{34}
\end{equation}
\[
\hspace{-0mm} 
 -\sum_{\epsilon_3\epsilon_4} \rule{0mm}{5mm}^{\prime} \,\, 
{\textstyle \frac{1}{2}} \epsilon_1\epsilon_2 \theta_{n_1}\theta_{n_2}\nu_{n_1-1}\nu_{n_2-1}  
\int_0^{\infty} \mbox{d}\omega |g(\omega)|^2\,  
\check{W}'_0 (z-\omega)_{(\epsilon_3,n_1-1)(\epsilon_4,n_2-1)} \, . 
\]
The definition
\begin{equation} 
\bar{W}'_0 (z)_{(\epsilon^{}_1,n^{}_1)(\epsilon^{}_2,n^{}_2)} =  
\left[ \check{W}'_0 (z^{\ast})_{(\epsilon^{}_1,n^{}_1)(\epsilon^{}_2,n^{}_2)}\right]^{\ast}  
\label{35} 
\end{equation} 
specifies the outcome of transforming the adjoint Kraus matrix. 

The atomic density matrix $\rho_a(t)$ must be computed from 
\begin{equation} 
\left._{a}\langle k|\right. \rho_a(t) 
|l\rangle_a = \sum_{n=0}^{\infty} \left._{a}\langle k|\right. \otimes \left._{f}\langle n|\right. 
\rho_S^{(1)}(t)\, |l\rangle_a \otimes |n\rangle_f \, .  
\label{36}  
\end{equation} 
If the matrix elements on the right-hand side are represented in terms of the states (\ref{28}), 
the iterative solution of (\ref{33}) can be exploited. For the probability that the atom is  
in the excited state at time $t$ we then arrive at the result  
\[ 
\hspace{-25mm} 
\left._{a}\langle 2|\right. \rho_a(t) |2\rangle_a = \sum_{r=0}^{\infty} 
\sum_{\{\epsilon^{}_s,\epsilon'_s\}_{s=1}^{2r+2}}^{\,\,\,\,\,\,\,\,\,\,\,\,\,\,\,\,\,\prime} 
\sum_{\{n^{}_s,n'_s=-1\}_{s=1}^{r+2}}^{\infty} \int_{-\infty}^{\infty} \frac{\mbox{d}\omega}{2\pi} 
\int_{-\infty}^{\infty} \frac{\mbox{d}\omega'}{2\pi}   
\int_0^{\infty}\mbox{d}\omega_1  \cdots \int_0^{\infty}\mbox{d}\omega_r  
\]
\[
\hspace{-25mm} 
\times \prod_{s=0}^r \left[
\check{W}'_0 (\omega+\omega^+_s+i0)_{(\epsilon^{}_{2s+1},n^{}_{s+1}+1)(\epsilon^{}_{2s+2},n^{}_{s+2})} 
\bar{W}'_0 (\omega'+\omega^+_s-i0)_{(\epsilon'_{2s+1},n'_{s+1}+1)(\epsilon'_{2s+2},n'_{s+2})} \right ] 
\] 
\[
\hspace{-25mm} 
\times \left( {\textstyle \frac{1}{2}} \right)^{r+1} \delta_{n^{}_1,n'_1} \epsilon_1 \epsilon'_1 \prod_{s=1}^r 
\left ( \epsilon^{}_{2s+1} \epsilon'_{2s+1} \nu_{n^{}_{s+1}}\nu_{n'_{s+1}} |g(\omega_s)|^2 \right )   
\langle \epsilon^{}_{2r+2},n^{}_{r+2}|\rho_S(0)| \epsilon'_{2r+2},n'_{r+2}\rangle
\]
\begin{equation}
\hspace{-25mm}
\times \, \exp (-i\omega t+i\omega' t+i\Omega_{\epsilon^{}_1,n^{}_1+1}t-i\Omega_{\epsilon'_1,n^{}_1+1}t) \, ,
\label{37}
\end{equation}  
where the abbreviation $\omega^+_s = \omega_1 + \omega_2 + \cdots + \omega_s$ is employed. Contours have been 
laid in accordance with the prescription derived in the previous section. The matrix element of the initial 
state $\rho_S(0)$ can be computed from (\ref{30}), whereas the matrices $\check{W}'_0$ and $\bar{W}'_0$ give rise 
to continued fractions after inversion and subsequent iteration of (\ref{34}). Obviously, the diagonal element 
$\left._{a}\langle 1|\right. \rho_a(t) |1\rangle_a $ follows from conservation of trace. Finally, the solution 
for the off-diagonal element $\left._{a}\langle 2|\right. \rho_a(t) |1\rangle_a $ can be found in the same manner 
as discussed above. 

In \cite{EPL:2013} it was argued that for large times $\rho_S(t)$ converges to the ground state. Such asymptotic 
behaviour gives rise to the limit  
\begin{equation}   
\lim_{t\rightarrow\infty} \rho_a(t) = \left ( \begin{array}{cc} 1 & 0 \\ 0 & 0 \end{array} \right )  \,\,\, ,   
\label{38}
\end{equation} 
the verification of which is indeed possible with the help of our solution for $\rho_a(t)$. To see how things 
work out, we recall the conclusions of the previous section on the analytic properties of the perturbative Kraus 
matrix. From this material it may be inferred that the matrix elements 
$\check{W}'_0 (\omega +i0)_{(\epsilon_1,n_1+1)(\epsilon_2,n_2)}$ are smooth and bounded. By an argument of 
Riemann-Lebesgue type one subsequently proves that the matrix elements 
$\left._{a}\langle 2|\right. \rho_a(t) |2\rangle_a$ and $\left._{a}\langle 2|\right. \rho_a(t) |1\rangle_a$ tend 
to zero for large times. The asymptotic behaviour of the other diagonal matrix element is determined by the Kraus 
matrix $\check{W}'_0 (\omega+i0)_{(+1,-1)(+1,-1)}$. It contains a pole at $\omega = \Omega_{+1,-1}$ which 
generates a contribution that does not depend on time. In view of trace conservation this contribution must be 
equal to one. 

The asymptotic decay of (\ref{37}) can be made more explicit upon taking the weak-coupling limit 
$g \rightarrow \lambda \tilde{g}$, $t \rightarrow \tilde{t}/\lambda^{\alpha}$, and $\lambda \rightarrow 0$, with 
$\alpha > 2$. By scaling in (\ref{37}) as $\omega \rightarrow \lambda^2 \tilde{\omega} + \Omega_{\epsilon_1,n_1+1}$ and 
$\omega' \rightarrow \lambda^2 \tilde{\omega}' + \Omega_{\epsilon'_1,n^{}_1+1}$ we eliminate fast Rabi oscillations 
from the exponential factor. Next, the forms   
$\lambda^2 \check{W}'_0(\lambda^2 \tilde{\omega}+\Omega_{\epsilon_1,n_1+1}+i0)_{(\epsilon_1,n_1+1)(\epsilon_2,n_2)}$ 
and $\lambda^2 \bar{W}'_0
(\lambda^2 \tilde{\omega}'+\Omega_{\epsilon'_1,n^{}_1+1}-i0)_{(\epsilon'_1,n^{}_1+1)(\epsilon'_2,n'_2)}$  
can be evaluated with the help of (\ref{35}) and the limit  
\[
\lim_{\lambda \rightarrow 0} \lambda^2 
\check{W}'_0(\lambda^2 \tilde{\omega}+ \Omega_{\epsilon_1,n_1}+i0)_{(\epsilon_1,n_1)(\epsilon_2,n_2)} = 
\] 
\begin{equation} 
\delta_{\epsilon_1,\epsilon_2}\delta_{n_1,n_2} \left [ \tilde{\omega} + \sum_{\epsilon}\rule{0mm}{5mm}^{\prime} 
{\textstyle \frac{1}{2}} \theta_{n_2} \nu^2_{n_2-1} \int_0^{\infty} \mbox{d} \omega   
\frac{|\tilde{g}(\omega )|^2}{\omega  + \Omega_{\epsilon,n_2-1} - \Omega_{\epsilon_2,n_2} - i0} \right]^{-1}  
\!\!\! ,   
\label{39}  
\end{equation} 
where the prime indicates that terms containing $\Omega_{-1,-1}$ must be excluded from the summation.  
One can verify the above result through iterating (\ref{34}) and dropping terms of order $\lambda^2$. 
Inverting the right-hand side of (\ref{34}) via the expansion $(A+B)^{-1} = A^{-1} - A^{-1} B A^{-1} + \cdots$, 
one recognizes that (\ref{39}) is indeed a diagonal matrix.    

For $n_2\ge 0$ and $\lambda$ small, the Kraus matrix (\ref{39}) generates exponential decay to zero in (\ref{37}). 
This is due to a pole lying below the real axis of the complex $\tilde{\omega}$ plane. To ascertain the location 
of this dissipative pole one should recognize that the difference $\Omega_{\epsilon_2,n_2} - 
\Omega_{\epsilon,n_2-1} = \omega_f + \epsilon_2 f (n_2+1)^{1/2} - \epsilon f (n_2)^{1/2}$ is positive for $n_2\ge 0$, 
$\epsilon = -1$, and $\omega_f > f $.  The afore-mentioned dissipative pole is also found upon taking the 
weak-coupling limit of the matrix element $\left._{a}\langle 2|\right. \rho_a(t) |1\rangle_a$. Therefore, in the 
weak-coupling regime the evolution of the atomic density matrix is in tune with the limit (\ref{38}).  

In Appendix B we demonstrate that for small coupling parameter $g$ and large initial photon number $p$ the evolution 
(\ref{37}) can be approximated as  
\begin{equation}  
\hspace{-12mm} 
\left._{a}\langle 2|\right. 
\rho_a( t ) |2 \rangle_a \simeq F (\tau) \equiv \rho_{a,22} \exp (-\tau )  
+ {\textstyle \frac{1}{2}} \exp (-\tau )   
\sum_{r=1}^p  ( 1- \rho_{a,11}\delta_{r,p} ) \frac{\tau^r}{r!}    
\,\,\, , 
\label{40} 
\end{equation}   
with the scaled time given by $\tau = \pi |g(\omega_f)|^2 t/2$. Surprisingly, for times of order 
$p/( \pi |g(\omega_f)|^{2})$ the existence of the limit (\ref{38}) is not reflected in the atomic evolution at all.  
Figure 1 shows that during the afore-mentioned time span the diagonal element (\ref{40}) takes on the value of 1/2.   
Only for times larger than $p/( \pi |g(\omega_f)|^{2})$ exponential decay to the ground state sets in. 

From (\ref{26A}) we see that the lifetime $\tau_e$ of the excited level of the free atom is equal to $1/(2\gamma)$. 
Since $\omega_f$ equals $\omega_{(21)}$ the plateau with $F (\tau) = 1/2$ thus appears at times of order $2 p\tau_e$. The 
initial photon number $p$ may be chosen arbitrarily large, so there is good reason to believe that in the presence of 
initial correlations between system and reservoir the behaviour depicted in Figure 1 will remain intact. This assertion 
is based on the expectation that assumption (ii) of the Introduction affects the evolution of the system up to a certain 
number of lifetimes $\tau_e$ and not at all times. 

\begin{figure}[h] 
\begin{center} 
\includegraphics[height=5cm]{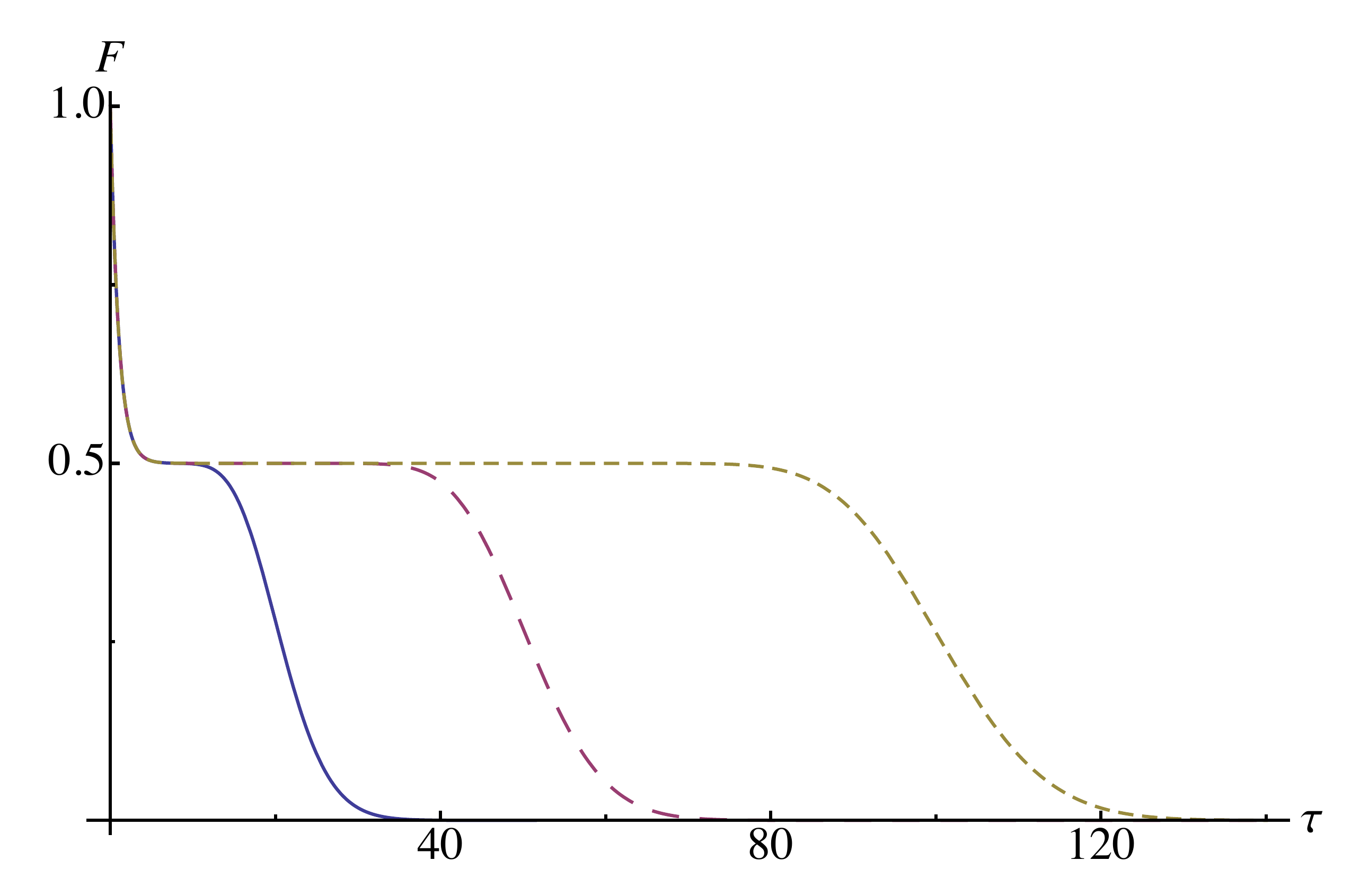} 
\end{center} 
\caption{Plot of evolution function $F ( \tau )$ defined in (\ref{40}) for $\rho_{a,22}=1$ and $p=20$  
(\rule[1mm]{12pt}{0.1pt}), $p=50$ (\rule[1mm]{6pt}{0.1pt} \rule[1mm]{6pt}{0.1pt} \rule[1mm]{6pt}{0.1pt}), 
and $p=100$ (\rule[1mm]{2pt}{0.1pt} \rule[1mm]{2pt}{0.1pt} \rule[1mm]{2pt}{0.1pt}). } 
\end{figure}

As the off-diagonal elements of the atomic density matrix are found to decay to zero for times of order  
$|g(\omega_f)|^{-2}$, the plateau of Figure 1 with $\left._{a}\langle 2|\right. \rho_a( t ) |2 \rangle_a = 1/2$ 
corresponds to the state of maximum entropy. The atomic propensity towards maximum entropy is 
corroborated by the limit 
\begin{equation}   
\lim_{ \lambda \rightarrow 0} 
\, \rho_a(g = \lambda \tilde{g}\, , p = \tilde{p}/\lambda^{\beta}\, , t = \tilde{t}/\lambda^{\alpha}) = 
\left ( \begin{array}{cc} 1/2   & 0 \\ 0 & 1/2 \end{array} \right )  \,\,\, ,  
\label{41} 
\end{equation}   
where the variables $\tilde{g}$, $\tilde{p}$, and $\tilde{t}$ must be kept constant, with 
$\tilde{p}/\lambda^{\beta}$ being of integer value. Furthermore, the conditions $2 < \alpha < \beta + 2$ and 
$0 < \beta < 4/3$ must be satisfied. A proof of the above limit is outlined in Appendix B. 

In this section, we have obtained a solution of non-Markovian character for the Jaynes-Cummings model with atomic 
damping by a transverse radiation field. Working at zero temperature, we have uncovered an analytical structure 
consisting of products of matrix continued fractions. Conservation of positivity and probability has been 
taken care of from the very outset. The non-Markovian evolution has been examined by performing analytical work, 
that is to say, by taking suitable limits. Both the ground state and the state of maximum entropy have been 
identified as attractor of the dissipative dynamics. 

\section{Conclusion} 

One of the main objectives of the theory of open quantum systems consists of describing, predicting and understanding 
experimental observations. A fundamental and well-known method to fulfill this ambition comprises the derivation and 
analysis of master equations for density operators. Unfortunately, it has become apparent, in particular from the 
recent literature \cite{CHR:2013, SEM:2016, FER:2017}, that the mathematics involved is very complicated, giving rise to 
virtually insurmountable technical barriers if one wishes to obtain exact non-Markovian results. For that reason, we 
decided some time ago to shift our attention from master equations to Kraus maps for evolutions of density matrices, 
opting for the system-reservoir setting as outlined in the Introduction.   

In the absence of initial correlations between system and reservoir, any dissipative evolution of a state $\rho_S$ in time 
$t$ is governed by the Kraus map $\sum_{j=1}^{\infty} W_j(t) \rho_S W_j^{\dagger}(t)$ \cite{ALI:2007}. Evaluation of 
the Kraus matrices $\{W_j(t)\}_j$ requires us to solve a time-ordering problem \cite{EPL:2013}, a most complex task 
\cite{DAT:1986}. Indeed, as it turns out in section 3 the exact solution for each Kraus matrix is given by an infinite  
continued fraction of a disturbingly difficult structure. This outcome most likely impedes any possibility of 
analytically performing the infinite summation figuring in the above Kraus map. We therefore corroborate conclusions 
from the recent articles cited above indicating that an exact description of dissipative dynamics by a finite set of 
evolution equations probably lies beyond our reach.         

In the Kraus map partial summation of terms becomes feasible as soon as one resorts to perturbation theory. This 
happens by carrying out factorizations such that the expansion for the density matrix is chopped up into bunches of $N$ 
terms. Owing to the self-similarity of the ensuing perturbative expansion, the possibility of deriving a finite set of 
evolution equations does exist now for arbitrary perturbative order. As shown in section 4, this set consists of a 
finite Kraus hierarchy as well as an integral equation for a bitemporal matrix. The latter gives us direct access to 
the perturbative density matrix, the positivity and trace of which are conserved in time. In lowest perturbative order, 
the dissipative dynamics depends on one Kraus matrix only. Its Laplace transform has neat analytical properties and can 
be represented by a much simpler continued fraction as compared to the non-perturbative case. As discussed in 
\cite{EPL:2013}, the exact density matrix and the perturbative density matrix coincide for large times and arbitrary 
order $N$.  

The work presented in section 3 demonstrates that for short times, when transients from the initial state and memory 
effects from the reservoir still have a large impact, exact evaluation of dissipative Kraus maps will demand excessive  
efforts. In fact, one may doubt whether the concept of reduced dynamics, i.e., embedding all equations of motion in the 
system's Hilbert space, is the optimal starting-point for studying early stages of dissipative dynamics. At the same time, 
our caveat does not imply that system-reservoir theory should be completely abandoned. On the contrary, outside the 
short-time regime the perturbative tools developed in section 4 can be put to use. This opens up the possibility of 
thoroughly examining non-Markovian evolutions towards thermal equilibrium by analytical means.   

To illustrate the last remark, we show in section 5 that the lowest-order perturbative density matrix of the 
Jaynes-Cummings model with non-Markovian radiative damping can be readily computed. In order to keep formulas as 
concise as possible, we assume zero temperature and discard counter-rotating contributions in the Hamiltonian. The 
evolution of the two-level atom can be analytically explored, namely by taking suitable asymptotic limits. We predict 
that the atom may remain in the state of maximum entropy for a significant time span that depends on the initial energy 
of the resonant radiation field as well as the energy loss to the transverse radiation field. 

Within the framework of Markovian dynamics, it has been shown \cite{JPA:2002} that the role of the state of maximum 
entropy as intermediate attractor occurs for a large class of initial states, including the case that the radiation 
field starts from a coherent state. Hence, experimental observation of the plateau depicted in Figure 1 might be 
feasible. A setup is required for which an atomic evolution can be monitored during a number of lifetimes of the 
excited level of the free atom that is of the same order as twice the initial average photon number of the radiation 
field.

\appendix
\setcounter{section}{0} 

\section{Conservation of probability}

We set out to prove that the derivative 
$ \sum_{k_1} \mbox{d} \langle k^{}_1 | \rho^{(N)}_S (t) | k_1 \rangle / \mbox{d} t $ 
as determined by (\ref{19}) is vanishing. The proof for the case $N=1$ was 
given in \cite{JPA:2006}, so we assume $N\ge 2$.  

We shall need partial derivatives of the Kraus matrices $\{W'_q \}_{q=0}^{N-1}$ 
with respect to $t$. From (\ref{8}) we obtain the identity 
\begin{equation} 
\hspace{-24mm} 
\frac{\partial}{\partial t} 
W'_q (t;T_q)_{K_{q+1}L_{q+1}}  =  
\label{A1}   
\end{equation} 
\[
\hspace{-24mm} 
- \sum_{j=1}^{q+1} \sum_{klm} \int_{t_j}^{t_{j-1}} \mbox{d} u  
\exp [i\omega_{(k_1k)}t]
W'_{q+1} (t;T^{}_{j-1},u,T^{j-1}_q)_{(kK^1_jmK^j_{q+1})(L^{}_{j-1}lL^{j-1}_{q+1})}  
c_{(k_1k)(lm)}(t,u)\,\,\,  ,
\]
where $t_0=t$ and $t_{q+1}=0$ must be substituted. Here and in the following, for $W'_N$ 
the factorized form  (\ref{17}) must be inserted.  

For the time being, we focus on the first and second contribution on the right-hand 
side of (\ref{19}). The time derivative that must be evaluated is given by 
\begin{equation}
D_1 = \frac{\partial}{\partial t} \sum_{k^{}_1 k'_1 l^{}_1 l'_1} W'^{}_0 (t)_{(k^{}_1)( l^{}_1)}\, 
\langle l^{}_1|\rho_S(0)| l'_1 \rangle \, W'^{\,\dagger}_0 (t)_{(l'_1)( k'_1)}  \delta_{k^{}_1k'_{1}}
\label{A2}  
\end{equation} 
\[
\hspace{11mm} 
+ \, \frac{\partial}{\partial t} 
\sum_{q=1}^{N-1} \sum_{k^{}_1 k'_1 l^{}_1 l'_1} \int_0^t 
\mbox{d}t^{}_1 \int_0^t \mbox{d} t'_1 \,\, 
\Delta'_q [\psi ] (t,t ;t^{}_1,t'_1;0,0)_{(k^{}_1 l^{}_1)(k'_1 l'_1)} 
\delta_{k^{}_1k'_{1}} \, .  
\]  
The first term on the right-hand side can be handled with the help of the choice $q=0$ in (\ref{A1}). 
In order to avoid any ambiguity in applying definition (\ref{5}), traces are taken by performing a sum 
over $k^{}_1$ and $k'_1$ with $\delta_{k^{}_1k'_{1}}$ as a weight. Defining  
\[
D_2 = \sum_{q=1}^{N-1} \sum_{k^{}_1 k'_1 l^{}_1 l'_1} \int_0^t 
\mbox{d}t^{}_1 \int_0^t \mbox{d} t'_1 \,\,  \frac{\partial}{\partial t} 
\Delta'_q [\psi ] (t,t ;t^{}_1,t'_1;0,0)_{(k^{}_1 l^{}_1)(k'_1 l'_1)} 
\delta_{k^{}_1k'_{1}} \, ,     
\] 
\[   
D_3 = \sum_{q=2}^{N-1} \sum_{k^{}_1 k'_1 l^{}_1 l'_1} \int_0^t \mbox{d} t'_1 \,\, 
\left. \Delta'_q [\psi ] (t,t ;t^{}_1,t'_1;0,0)_{(k^{}_1 l^{}_1)(k'_1 l'_1)}\right|_{t_1 = t} 
\,\, \delta_{k^{}_1k'_{1}} \, ,  
\]  
\begin{equation}
D_4 = \sum_{q=2}^{N-1} \sum_{k^{}_1 k'_1 l^{}_1 l'_1} \int_0^t \mbox{d} t_1 \,\, 
\left. \Delta'_q [\psi ] (t,t ;t^{}_1,t'_1;0,0)_{(k^{}_1 l^{}_1)(k'_1 l'_1)}\right|_{t'_1 = t} 
\,\, \delta_{k^{}_1k'_{1}} \,\,\, , 
\label{A3} 
\end{equation} 
we can express the result for  the time derivative (\ref{A2}) as $D_1 = D_2 + D_3 + D_4$. 

Upon employing (\ref{A1}) in $D_2$ one obtains 
\[ 
D_2 = - \sum_{q=1}^{N-1} \sum_{K^{}_{q+2}L^{}_{q+2}K'_{q+2}L'_{q+2}} 
\int_0^t \mbox{d} T^{}_{q+1} \int_0^t \mbox{d} {T'}^1_{q+1}  
\]
\[
\hspace{11mm} 
\times W'_{q+1} (t;T^{}_{q+1})_{K^{}_{q+2}L^{}_{q+2}}  \psi(0,0)_{(l^{}_{q+2})(l'_{q+2})}
 W'^{\,\dagger}_{q+1} (t;t, {T'}^{1}_{q+1})_{L'_{q+2}K'_{q+2}}  
\]
\begin{equation} 
\hspace{11mm} 
\times \sum_{PQ} \sum_{j=1}^{q+1} \frac{1}{q!} \prod_{h=1}^q \{\tilde{Q}(h)'\,\,\tilde{P}(h)\} 
\left. \{ 1' \,\, j \}\right|_{t'_1 = t} \delta_{k^{}_1k'_{1}} + \mbox{c.c.}\,\,\, ,
\label{A4}
\end{equation} 
where one must substitute $\tilde{P}(h)=P(h)+\sum_{i=j}^q 
\delta_{P(h),i}$ and $\tilde{Q}(h)=Q(h)+1$. 
The symbols $P$ and $Q$ denote permutations of the integers 
$\{1,2,\ldots,q\}$. Variables have been transformed as $u\rightarrow t_j$, $m\rightarrow k_{j+1}$, 
$l\rightarrow l_j$, $t^{}_i\rightarrow t^{}_{i+1}, k^{}_{i+1}\rightarrow k^{}_{i+2}, 
l^{}_i\rightarrow l^{}_{i+1}$ for $i=j,j+1,\ldots,q$ and $t'_i\rightarrow t'_{i+1}, 
k'_{i+1}\rightarrow k'_{i+2}, l'_i\rightarrow l'_{i+1}$ for $i=1,2,\ldots,q$. Also, the shifts 
$k\rightarrow l'^{}_1$, $l_{q+1}\rightarrow l_{q+2}$, and $l'_{q+1}\rightarrow l'_{q+2}$ have been 
performed. Identity (\ref{8}) has been applied to $W'^{\,\dagger}_{q+1}$, with $t'_1$ set equal to $t$.     

In (\ref{A4}) a combinatorial sum over products of correlation functions is performed. Employing  
Kronecker deltas we extend the permutations $P$ en $Q$ with the elements $P(q+1)$ and $Q(q+1)$, 
respectively. The ensuing expression can be simplified by means of the identity   
\[
\sum_{PQ} \sum_{j=1}^{q+1} \prod_{h=1}^q \{ \tilde{Q}(h)' \,\,  \tilde{P}(h) \} \{ 1' \,\,  j \} 
\delta_{Q(q+1),q+1} \delta_{P(q+1),q+1} = 
\]
\begin{equation}  
\frac{1}{q+1} \sum_{PQ} \prod_{h=1}^{q+1}  \{ Q(h)' \,\, P(h) \} \,\,\, , 
\label{A5} 
\end{equation} 
which can be verified by inspection. On both sides of (\ref{A5}) we sum over all permutations 
$P$ and $Q$ of the integers $\{1,2,\ldots,q+1\}$. 

Owing to (\ref{A5}) we can phrase (\ref{A4}) in the following concise manner  
\begin{equation} 
\hspace{-14mm} 
D_2 = - \sum_{q=2}^{N} \sum_{k^{}_1 k'_1 l^{}_1 l'_1} \int_0^t \mbox{d} t_1 \,\, 
\left. \Delta'_q [\psi ] (t,t ;t^{}_1,t'_1;0,0)_{(k^{}_1 l^{}_1)(k'_1 l'_1)}\right|_{t'_1 = t} 
\,\, \delta_{k^{}_1k'_{1}} + \mbox{c.c.} \,\,\, . 
\label{A6} 
\end{equation}  
Note that first one should perform all permutations contained in $\Delta'_q$ and then make 
the substitution $t'_1=t$. In the c.c. term one should make the substitution $t_1=t$. 
 
Except for the term with $q=N$ the right-hand side of (\ref{A6}) cancels out against the sum 
of $D_3$ and $D_4$. Therefore, the desired derivative $D_1$ attains the form 
\begin{equation} 
\hspace{-14mm} 
D_1 = -  \sum_{k^{}_1 k'_1 l^{}_1 l'_1} \int_0^t \mbox{d} t_1 \,\, 
\left. \Delta'_N [\psi ] (t,t ;t^{}_1,t'_1;0,0)_{(k^{}_1 l^{}_1)(k'_1 l'_1)}\right|_{t'_1 = t} 
\,\, \delta_{k^{}_1k'_{1}} + \mbox{c.c.} \,\,\, . 
\label{A7} 
\end{equation}  
In making $\Delta'_N$ explicit one should employ the truncation prescription (\ref{17}). 

The derivatives of the third and fourth term on the right-hand side of (\ref{19}) can be treated 
in a similar manner as explained above. Abbreviating these contributions as 
\[ 
\hspace{0mm} 
D_5 =  
\frac{\partial}{\partial t} 
\sum_{k^{}_1 k'_1 l^{}_1 l'_1} \int_0^t \mbox{d}s \int_0^t \mbox{d} s' 
\exp [i\omega_{k^{}_1}s - i\omega_{k'_1}s']   
\]
\[ 
\hspace{11mm} 
\times W'^{}_0 (t-s)_{(k^{}_1)( l^{}_1)}\, \chi^{(N)} (s,s')_{(l^{}_1)( l'_1)}\, 
W'^{\,\dagger}_0 (t-s')_{(l'_1)( k'_1)}  \delta_{k^{}_1k'_{1}} 
\]
\[
\hspace{11mm} 
+ \frac{\partial}{\partial t}  
\sum_{q=1}^{N-1} \sum_{k^{}_1 k'_1 l^{}_1 l'_1} \int_0^t \mbox{d}s \int_0^t \mbox{d} s' 
\int_s^t \mbox{d} t^{}_1 \int_{s'}^t \mbox{d} t'_1 \exp [i\omega_{k^{}_1}s - i\omega_{k'_1}s']   
\] 
\begin{equation} 
\hspace{11mm} 
\times \Delta'_q [\chi^{(N)} ] (t,t ;t^{}_1,t'_1;s,s')_{(k^{}_1 l^{}_1)(k'_1 l'_1)}   
\delta_{k^{}_1k'_{1}}  \, , 
\label{A8}  
\end{equation} 
we obtain 
\[
\hspace{-0mm} 
D_5 =
\sum_{l^{}_1 l'_1} \int_0^t \mbox{d} s' 
\exp [i\omega_{l^{}_1}(t - s')] \chi^{(N)} (t,s')_{(l^{}_1)( l'_1)}\, 
W'^{\,\dagger}_0 (t-s')_{(l'_1)( l^{}_1)}  
\] 
\[ 
\hspace{11mm}   
- \sum_{k^{}_1 k'_1 l^{}_1 l'_1} \int_0^t \mbox{d}s \int_0^t \mbox{d} s' 
\int_s^t \mbox{d} t^{}_1 \exp [i\omega_{k^{}_1}s - i\omega_{k'_1}s']   
\] 
\begin{equation}
\hspace{11mm} 
\times \left. \Delta'_N [\chi^{(N)} ] (t,t ;t^{}_1,t'_1;s,s')_{(k^{}_1 l^{}_1)(k'_1 l'_1)} 
\right|_{t'_1 = t} \,\,   
\delta_{k^{}_1k'_{1}} + \mbox{c.c.} \,\,\, .     
\label{A9}
\end{equation} 
The complex conjugate of the two terms on the right-hand side must be added and in 
the second complex conjugate $t_1=t$ must be chosen. 

Upon combining definition (\ref{5}) with prescription (\ref{17}) we see that on the right-hand side of (\ref{A9}) 
evolution equation (\ref{20}) can be invoked. Then (\ref{A9}) simplifies to the form 
\begin{equation}
\hspace{-6mm} 
D_5 = 
\sum_{k^{}_1 k'_1 l^{}_1 l'_1} \int_0^t \mbox{d} t_1 \,\, 
\left. \Delta'_N [\psi ] (t,t ;t^{}_1,t'_1;0,0)_{(k^{}_1 l^{}_1)(k'_1 l'_1)}\right|_{t'_1 = t} 
\,\, \delta_{k^{}_1k'_{1}} + \mbox{c.c.} \,\,\, . 
\label{A10} 
\end{equation} 
Addition of (\ref{A7}) and (\ref{A10}) yields zero. This completes the proof.   

\section{Limit of maximum entropy} 

In demonstrating the validity of the limit (\ref{41}) we shall interchange limits and infinite sums without proof. 
A Markovian counterpart of (\ref{41}) was rigorously proved in \cite{JPA:2002}.     

By performing the scaling $g \rightarrow \lambda \tilde{g}$, $p \rightarrow \tilde{p}/\lambda^{\beta}$, 
$t \rightarrow \tilde{t}/\lambda^{\alpha}$, $\omega \rightarrow \lambda^2 \tilde{\omega} + \Omega_{\epsilon_1,n_1+1}$,  
and $\omega' \rightarrow \lambda^2 \tilde{\omega}' + \Omega_{\epsilon'_1,n^{}_1+1}$, with $\alpha$, $\beta > 0$, we 
eliminate fast Rabi oscillations from the exponential factor of (\ref{37}) and pave the way for use of (\ref{39}). With 
the help of the identity 
\begin{equation} 
\Omega_{\epsilon,n-1} - \Omega_{\epsilon',n} = - \omega_f  + \epsilon f (n)^{1/2}  
- \epsilon' f (n+1)^{1/2}   
\label{B1} 
\end{equation}  
the denominator of (\ref{39}) can be elaborated. 

Upon inserting (\ref{30}) into (\ref{37}) one recognizes that integer $n$ is of order $\tilde{p}/\lambda^{\beta}$. 
For the choice $\epsilon = \epsilon'$ the right-hand side of (\ref{B1}) converges to $-\omega_f$ as $\lambda$ becomes 
small. In contrast, the choice $\epsilon = - \epsilon'$ makes a contribution to (\ref{39}) that decays as $\lambda^{\beta /2}$. 
For large $n$ and $n'$ we thus arrive at 
\begin{equation} 
\hspace{-10mm} 
\lim_{\lambda \rightarrow 0} \lambda^2 
\check{W}'_0(\lambda^2 \tilde{\omega}+\Omega_{\epsilon,n}+i0)_{(\epsilon,n)(\epsilon',n')} = 
\delta_{\epsilon,\epsilon'}\delta_{n,n'} 
\left ( \tilde{\omega} + i\Gamma + \bar{\omega}_f \right )^{-1} \,\,\, ,  
\label{B2} 
\end{equation}    
with $\Gamma$ and $\bar{\omega}_f$ given by $\pi |\tilde{g}(\omega_f)|^2/4$ and 
$(1/4){\cal P}\int_0^{\infty} \mbox{d}\omega |\tilde{g}(\omega)|^2/(\omega - \omega_f)$, respectively. 

Transforming integral dummies as $\omega^+_s \rightarrow \lambda^2 \tilde{\omega}_s + 
\Omega_{\epsilon_{2s+1},n_{s+1}+1}-\Omega_{\epsilon_1,n_1+1}$ for $1\le s \le r$, we encounter the forms  
$\lambda^2 \check{W}'_0(\lambda^2 \tilde{\omega} + \lambda^2 \tilde{\omega}_s 
+ \Omega_{\epsilon_{2s+1},n_{s+1}+1}+i0)_{(\epsilon_{2s+1},n_{s+1}+1)(\epsilon_{2s+2},n_{s+2})}$, with 
$0\le s \le r$ and $\tilde{\omega}_0 = 0$. For small $\lambda$, these can be computed on the basis of (\ref{B2}), 
whereafter the conditions $\epsilon_{2s+1} = \epsilon_{2s+2}$ and $n_{s+1} +1 = n_{s+2}$ appear, with $0\le s \le r$.  

By (\ref{35}) and (\ref{B2}) the matrix $\lambda^2 \bar{W}'_0( \lambda^2 \tilde{\omega}' + 
\Omega_{\epsilon'_{1},n_{1}+1} -i0)_{(\epsilon'_{1},n_{1}+1)(\epsilon'_{2},n'_{2})}$ gives rise to the conditions 
$\epsilon'_{1} = \epsilon'_{2}$ and $n_{1} +1 = n'_{2}$, with $\lambda$ small. Hence, the $r=0$ term of (\ref{37}) 
converges to $\rho_{a,22} \exp [-2\Gamma \tilde{t} \lambda^{2-\alpha} ]$. For $r>0$ we meet matrices $\bar{W}'_0$ 
containing the sum 
\begin{equation} 
\Phi_s = \Omega_{\epsilon'_1,{n^{}_1+1}}  -  \Omega_{\epsilon^{}_1,n^{}_1+1}  
 - \Omega_{\epsilon'_{2s+1},{n'_{s+1}+1}} + \Omega_{\epsilon^{}_{2s+1},n^{}_{s+1}+1}  \,\,\, ,   
\label{B3} 
\end{equation} 
with $1\le s \le r$. As long as $\Phi_s$ decays slower than $\lambda^2$ for $\lambda\rightarrow 0$, the limit 
\begin{equation} 
\hspace{-18mm} 
\lim_{\lambda\rightarrow 0} \lambda^2 
\bar{W}'_0(\Phi_s + \lambda^2 \tilde{\omega}' + \lambda^2 \tilde{\omega}_s + 
\Omega_{\epsilon'_{2s+1},n'_{s+1}+1} -i0)_{(\epsilon'_{2s+1},n'_{s+1}+1)(\epsilon'_{2s+2},n'_{s+2})} = 0 \,\,\, 
\label{B4} 
\end{equation} 
is true. Hence, finite contributions to (\ref{37}) arise only if $\Phi_s$ decays faster than $\lambda^2$.  

Obviously, the condition $\Phi_s = 0$ is sufficient. As $\omega_f$ and $f$ are independent, it is 
equivalent to $n^{}_{s+1} = n'_{s+1}$ and 
$(\epsilon^{}_1 - \epsilon'_1)(n_1 + 2)^{1/2} = (\epsilon^{}_{2s+1} - \epsilon'_{2s+1})(n_{s+1}+2)^{1/2}$, 
with $1\le s \le r$. For $\epsilon^{}_1 \ne \epsilon'_1$ the last relation gives $n_1 = n_{s+1}$, an  
outcome that is contradictory to the identity $n_1+1 = n_2$ derived earlier. Hence, for $1\le s \le r$ we find 
$n^{}_{s+1} = n'_{s+1}$ and $\epsilon^{}_{2s+1} = \epsilon'_{2s+1}$. One verifies that for $\beta < 4$ these choices 
are not only sufficient but necessary as well. Once $\Phi_s$ has disappeared, the Kronecker deltas of (\ref{B2}) 
give rise to the equalities $\epsilon'_{2s+1} = \epsilon'_{2s+2}$ and $n'_{s+1} +1 = n'_{s+2}$, with $1\le s \le r$. 

Using (\ref{30}) we find that for $r>0$ the initial matrix element of (\ref{37}) reduces to        
\begin{equation} 
\langle \epsilon^{}_{2r+2},n^{}_{r+2}|\rho_S(0)| \epsilon'_{2r+2},n'_{r+2}\rangle =  {\textstyle \frac{1}{2}}   
(\rho_{a,11}\delta_{n_{r+2},p-1} +  \rho_{a,22} \delta_{n_{r+2},p}  )\,\,\, .    
\label{B5} 
\end{equation} 
For $1\le s \le r+2$ and $\rho_{a,22}=1$ we must choose $n^{}_s = n'_s = p + s - r - 2$ , so that the condition $r\le p$ 
emerges. Of course, for $\rho_{a,11}=1$ the replacement $p\rightarrow p-1$ must be carried out in $n^{}_s$ and $n'_s$. 
For small $\lambda$ the coupling constants of (\ref{37}) attain the form   
\begin{equation} 
\hspace{-0mm}
\tilde{g}(\omega_s) = \tilde{g}(  
\omega_f - \epsilon_{2s-1} f [n_1 + s + 1]^{1/2} + \epsilon_{2s+1} f [n_1 + s + 2]^{1/2} ) 
\,\,\, .  
\label{B6} 
\end{equation} 
Since $n_1$ is of order $\tilde{p}/\lambda^{\beta}$, the argument of $\tilde{g}$ 
diverges as $\lambda^{-\beta /2}$. From the assumption $\tilde{g}(\omega)\rightarrow 0$ for 
$\omega\rightarrow \infty$ it thus follows that the choice $\epsilon_{2s-1} = \epsilon_{2s+1}$ is compulsory. 
Then for all coupling constants the finite value of $\tilde{g}(\omega_f)$ is found. 

For the integrals over $\{\tilde{\omega}_s\}_{s=1}^r$ the domain of integration is determined by 
\begin{equation} 
\lambda^2 \tilde{\omega}_s > \lambda^2 \tilde{\omega}_{s-1} - \Omega_{\epsilon_{2s+1},n_{s+1}+1} 
+ \Omega_{\epsilon_{2s-1},n_s+1} \,\,\,  ,         
\label{B7} 
\end{equation} 
with $1\le s\le r$ and $\tilde{\omega}_0 =0$. For $\lambda$ tending to zero (\ref{B7}) boils down to 
$\tilde{\omega}_s > \tilde{\omega}_{s-1} - \omega_f /\lambda^2$ or $\tilde{\omega}_s > -\infty$. Hence, all 
integrals over $\tilde{\omega}_s$ can be computed with the help of the residue theorem. For the remaining 
integrals over $\tilde{\omega}$ and $\tilde{\omega}'$ the transformation $x = \tilde{\omega} - \tilde{\omega}'$ 
can be performed. If $\lambda$ is small this brings us to   
\[ 
\hspace{-0mm} 
\left._{a}\langle 2|\right. 
\rho_a(g = \lambda \tilde{g}\, , p = \tilde{p}/\lambda^{\beta}\, , t = \tilde{t}/\lambda^{\alpha}) 
|2 \rangle_a \rightarrow  \rho_{a,22} \exp [-2\Gamma \tilde{t} \lambda^{2-\alpha} ] 
\]
\begin{equation}  
\hspace{-0mm} 
+ \frac{i}{4\pi} 
\sum_{r=1}^p  ( 1- \rho_{a,11}\delta_{r,p} )
\int_{-\infty}^{\infty} \mbox{d} x \exp [-ix\tilde{t} \lambda^{2-\alpha} ] 
\frac{(2i\Gamma)^r}{(x+2i\Gamma)^{r+1}}
\,\,\, .
\label{B8} 
\end{equation}   
For $\alpha > 2$ use of the residue theorem leads to the evolution (\ref{40}) appearing in the main text. 
Upon combining a standard formula \cite{ABR:1965} with the condition 
$\alpha < \beta +2$ we obtain from (\ref{B8}) the desired result of $1/2$ for $\lambda$ small. 

Performing the same scaling as above, one shows that $\left._{a}\langle 2|\right. \rho_a(t) |1\rangle_a$ converges 
to zero in the limit (\ref{41}). In (\ref{B4}) the sum $\Phi_s$ must be exchanged for   
\begin{equation} 
\Phi'_s = \Omega_{\epsilon'_1,{n^{}_1+1}}  -  \Omega_{\epsilon^{}_1,n^{}_1}  
 - \Omega_{\epsilon'_{2s+1},{n'_{s+1}+1}} + \Omega_{\epsilon^{}_{2s+1},n^{}_{s+1}+1}  \,\,\, ,       
\label{B9} 
\end{equation}  
with $1\le s \le r$. All Kraus matrices $\bar{W}'_0$ figuring in $\left._{a}\langle 2|\right. \rho_a(t) |1\rangle_a$ 
tend to zero if $\Phi'_s$ decays slower than $\lambda^2$ for $\lambda\rightarrow 0$. From (\ref{28}) one obtains 
$\Phi'_s = (\epsilon'_1-\epsilon^{}_1) {\cal O} (\lambda^{\beta/2})+  {\cal O} (\lambda^{3\beta/2})$. Under the 
choice $\epsilon'_1 = \epsilon^{}_1$ the condition $\beta < 4/3$ is found.

\end{document}